%% file: SNO+_exp_AHEP_nmo15_online_VLozza_v5.tex
\documentclass[10pt,twocolumn,twoside]{article}
\usepackage[top=1in, bottom=1.8cm, outer=1.8cm, inner=1.8cm, headsep=14pt]{geometry}
\usepackage{graphicx}
\usepackage{subfigure}
\usepackage{tablefootnote}
\usepackage{multirow}
\usepackage{amsmath}
\usepackage{epsfig, float}
\usepackage{array, tabularx}
\usepackage{caption}
\usepackage{url}
\usepackage{titlesec}
\titlelabel{\thetitle. }
\usepackage{gensymb}
\usepackage{footnote}
\usepackage{tikz}
\usepackage{pgfplots}
\usepackage[noblocks]{authblk}
\usepackage{afterpage}
\usepackage{longtable}

\newcommand{\bnel}{\mbox{$\bar{\nu}_e$}}

\newcommand{\nel}{\mbox{$\nu_e$}}

\newcommand{\nx}{\mbox{$\nu_x$}} 
\newcommand{\vp}{\mbox{$\nu$--$p$ ES}} 

\newcommand{\obb}{0\mbox{$\nu\beta\beta$-decay} }
\newcommand{\zbb}{2\mbox{$\nu\beta\beta$-decay} }

\titleformat{\subsection}[runin]{\normalfont\fontsize{10}{10}\itshape}{\thesubsection.\noindent}{1em}{}[.]
\titleformat{\subsubsection}[runin]{\normalfont\fontsize{10}{10}\itshape}{\thesubsubsection.\noindent}{1em}{}[.]
\makeatletter
\newcommand\footnoteref[1]{\protected@xdef\@thefnmark{\ref{#1}}\@footnotemark}
\makeatother

\usepackage{fancyhdr}
\makeatletter
\patchcmd{\@maketitle}{center}{flushleft}{}{}
\patchcmd{\@maketitle}{center}{flushleft}{}{}

\def\maketitle{{%
  \renewenvironment{tabular}[2][]
    {\begin{flushleft}}
    {\end{flushleft}}
  \AB@maketitle}}
\makeatother

\begin{document} 

\begin{titlepage}
\title{
\begin{flushleft}
\textbf{Current Status and Future Prospects of\\ the SNO+ Experiment}
\end{flushleft}
}
%\date{\today}
\date{\vspace{-5ex}}
\input{201506_review_authorlist_v9b.tex}
%\hsize=\textwidth\linewidth=\textwidth
\maketitle
\thispagestyle{empty}
\end{titlepage}

\twocolumn[

%\title{\Large
%\begin{flushleft}
%\textbf{Current Status and Future Prospects of\\ the SNO+ Experiment -- Draft 4}
%\end{flushleft}
%}  

  \begin{@twocolumnfalse}

%begin{abstract}
%\label{abstract}

\hsize=\textwidth\linewidth=\textwidth
\small
\quotation\noindent 

\noindent SNO+ is a large liquid scintillator-based experiment located 2\,km underground at SNOLAB, Sudbury, Canada. It reuses the Sudbury Neutrino Observatory detector, consisting of a 12\,m diameter acrylic vessel which will be filled with about 780 tonnes of ultra-pure liquid scintillator. Designed as a multipurpose neutrino experiment, the primary goal of SNO+ is a search for the neutrinoless double-beta decay (0$\nu\beta\beta$) of $^{130}$Te. In Phase I, the detector will be loaded with 0.3\% natural tellurium, corresponding to nearly 800\,kg of $^{130}$Te, with an expected effective Majorana neutrino mass sensitivity in the region of 55-133\,meV, just above the inverted mass hierarchy. Recently, the possibility of deploying up to ten times more natural tellurium has been investigated, which would enable SNO+ to achieve sensitivity deep into the parameter space for the inverted neutrino mass hierarchy in the future. Additionally, SNO+ aims to measure reactor antineutrino oscillations, low energy solar neutrinos, and geoneutrinos, to be sensitive to supernova neutrinos, and to search for exotic physics. A first phase with the detector filled with water will begin soon, with the scintillator phase expected to start after a few months of water data taking. The $0\nu\beta\beta$ Phase I is foreseen for 2017.

%\end{abstract}

  \end{@twocolumnfalse}
  \vskip2.0em
  
]

\section{Introduction}\label{Intro}

SNO+ is a large-scale liquid scintillator experiment located at a depth of 5890 $\pm$ 94\,meter water equivalent (m.w.e.) in Vale's Creighton mine in Sudbury, Canada. The deep underground location, the high purity of materials used, and the large volume make SNO+ an ideally suited detector to study several aspects of neutrino physics. 

The main goal of SNO+ is a search for the neutrinoless double-beta decay (0$\nu\beta\beta$) of $^{130}$Te. \obb is a rare nuclear process that will happen if neutrinos are Majorana-type particles, that is, they are their own antiparticles. Understanding the Majorana nature of neutrinos is one of the most active areas of research in modern neutrino physics. The observation of the \obb would demonstrate lepton number violation, a key ingredient in the theory of leptogenesis. The process can be seen as two simultaneous $\beta$-decays, in which two neutrons are converted into two protons and two electrons, as the neutrinos from the two weak vertices mutually annihilate. The signature is a peak at the Q-value of the process in the summed energy spectrum of the two electrons. The measured quantity is the half-life of the decay. The effective Majorana neutrino mass, $m_{\beta\beta}$, which is highly dependent on the nuclear matrix elements, is derived from the half-life as described in \cite{zub04}. A half-life of the order of 10$^{25}$ years corresponds to a neutrino mass range of about 200--400\,meV. The large mass and low background of SNO+ allow the investigation of such a rare event.

The large volume and the high radio-purity are also the reason why SNO+ can explore several other physics topics. Observation of geo-neutrinos will help in understanding the mechanisms for heat production in the Earth. Reactor antineutrino measurements constrain the neutrino oscillation parameters. Neutrinos and antineutrinos coming from supernova explosions would help to answer many unresolved questions in neutrino astronomy. Additionally, SNO+ has the potential to search for exotic physics like axion-like particles and invisible nucleon decay.

The depth of SNOLAB also provides the opportunity to measure low energy solar neutrinos, like \textit{pep} and CNO neutrinos. The \textit{pep} neutrinos are monoenergetic, with an energy of 1.44\,MeV and a very well predicted flux, with an uncertainty of 1.2\%, constrained by the solar luminosity \cite{sere11}. A precise measurement of the flux can probe the Mikheyev, Smirnov and Wolfenstein (MSW) effect of neutrino mixing as well as alternate models like Non Standard Interactions \cite{pena}. Another open question in the solar neutrino field is related to the solar metallicity. The Standard Solar Model was always in excellent agreement with helioseismology until recent analyses suggested a metallicity about 30\% lower than the previous model. This raised the question of the homogeneous distribution of elements heavier than helium in the Sun. The measurement of the CNO neutrino flux could be used to solve the problem \cite{Serenelli}.\\

This paper is structured as follows. In Sections \ref{sec::SNO+} and \ref{sec::status}  the SNO+ experiment is described, including the current status and detector upgrades. The expected background sources are presented in Section \ref{sec::Background}. In Sections \ref{sec::doublebeta} to \ref{sec::exotic} the broad physics program of SNO+ is described: the neutrinoless double-beta decay search (Section \ref{sec::doublebeta}), the measurement of low energy solar neutrinos (Section \ref{sec::solar}), the measurements of geo and reactor antineutrinos (Section \ref{sec::antinu}), the supernova neutrino watch (Section \ref{sec::supernova}), and the exotic physics searches (Section \ref{sec::exotic}). A brief conclusion follows at the end.

\section{The SNO+ Experiment} \label{sec::SNO+}

\begin{figure}[t]\centering
\includegraphics[scale=0.45]{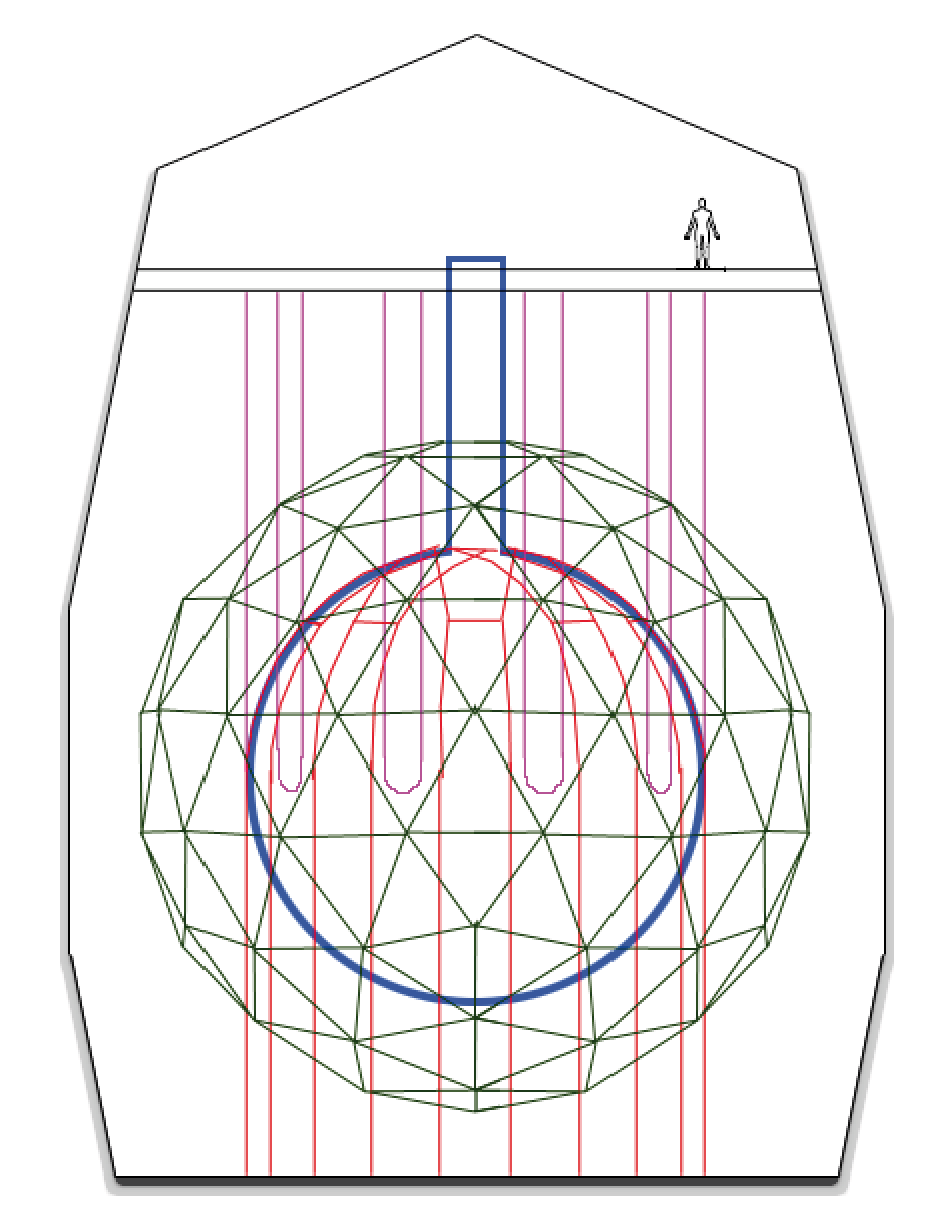} 
\caption{The SNO+ detector, figure from \cite{PJ_PhD}. The 12\,-m diameter acrylic vessel (blue) is viewed by $\sim$9300 PMTs supported by a $\sim$18\,-m diameter geodesic structure (green) and is held by a system of high purity ropes (purple). The AV and the PSUP are within a volume of highly purified water. A rope net (red) will be used to offset the buoyancy of the liquid scintillator contained within the AV. \label{fig::SNO_detector}}
\end{figure}

The SNO+ experiment \cite{chen04} is located in the underground laboratory of SNOLAB, Sudbury, Canada. A flat overburden of 2092\,m of rock provides an efficient shield against cosmic muons corresponding to 5890 $\pm$ 94\,m.w.e. \cite{sno09}. The resulting muon rate through a 8.3\,m radius circular area is 63 muons per day. SNO+ will make use of the SNO detector structure \cite{sno00, Jel09} consisting of a spherical acrylic vessel (AV) of 6\,m radius and 5.5\,cm thickness located within a cavity excavated in the rock. The vessel will be filled with about 780 tonnes of liquid scintillator and will be viewed by $\sim$9300 PMTs supported by a geodesic stainless steel structure (PSUP) of approximately 8.9\,m radius. The volume between the AV and the PSUP, as well as the rest of the cavity, will be filled with about 7000 tonnes of ultra-pure water, which acts as a shield for the radioactivity coming from the rock (cavity walls) and the PMT array. A system of hold-up ropes suspends the acrylic vessel inside the PSUP. Additionally, in order to balance the buoyant force due to the lower density of the liquid scintillator compared to the external water, a new system of hold-down ropes has been installed on the top part of the AV and anchored at the cavity floor. A sketch of the detector is shown in Figure\,\ref{fig::SNO_detector}. 

The major detector upgrades, including the liquid scintillator process systems, are described here.

\subsection{Liquid Scintillator}\label{LS}
The SNO+ liquid scintillator (LS) is composed of an aromatic hydrocarbon, linear alkylbenzene (LAB), as a solvent, and a concentration of 2\,g/L 2,5-diphenyloxazole (PPO) as a fluor. LAB was selected as the liquid scintillator for SNO+ because of (1) its long time stability, (2) compatibility with the acrylic, (3) high purity levels directly from the manufacturer, (4) long attenuation and scattering length, (5) high light yield, and (6) linear response in energy. Additionally, it has a high flash point and is environmentally safe. LAB will be produced very close to the detector location (at the Cepsa plant in Bécancour, Quebec, less than 900\,km away), allowing short transport times which are important to reduce the possibility of cosmogenic activation.

\subsection{Te-Loading}\label{sec::teloading}
One of the main advantages of using LAB as liquid scintillator is the possibility of dissolving heavy metals with long term stability and good optical properties. For the \obb phase of the experiment, SNO+ will load tellurium into the liquid scintillator. An innovative technique has been developed to load tellurium at concentration levels of several percent into LAB maintaining good optical properties and reasonably high light emission levels \cite{bnl11}. Telluric acid, Te(OH)$_{6}$, is first dissolved in water and then, adding a surfactant, loaded into the scintillator. To better match the PMT quantum efficiency a secondary wavelength shifter will also be added to the mixture. Currently, we are investigating two different secondary wavelength shifters: perylene and bis-MSB. The former shifts the emission peak's range from 350--380\,nm to $\sim$450--480\,nm with a predicted light yield in SNO+ of about 300\,Nhits (detected photoelectron hits) per MeV of energy. The latter shifts the emission peak to $\sim$390--430\,nm with a light yield of 200\,Nhits/MeV. The final choice will depend on the timing optical properties, the light yield, and the scattering length of the full scintillator mixture.

\subsection{Emission Timing Profiles and Optical Properties}\label{sec::time_Q}

The emission timing profile and the optical properties of the LAB-PPO and the Te-loaded scintillator have been thoroughly investigated. The timing profile of scintillation pulses depends on the ionization density of the charged particles, with signals caused by electrons being faster than those from protons or alpha particles. This property allows the discrimination among particle types, which is very important for background rejection. The timing profile of electron and alpha particles in the unloaded scintillator has been measured in \cite{chen11}. Results show that, for a LAB-PPO sample, a peak-to-total ratio analysis allows us to reject $>$ 99.9\% of the alpha particles while retaining $>$ 99.9\% of the electron signal. 

The measurement of the timing profiles in the 0.3\% Te-loaded scintillator is described in \cite{sean14}. The presence of water and the surfactant in the cocktail reduces the long tail of the alpha decay (slow component) with respect to the unloaded scintillator, resulting in a poorer discrimination between $\alpha$-like and $\beta$-like signals. %Preliminary studies show that to achieve 90\% alpha rejection nearly 20\% of the beta-like signals are sacrificed. 

The light yield of the unloaded LAB-PPO scintillator has been measured in bench top tests and extrapolated for the full SNO+ volume using Monte Carlo (MC) simulations, leading to 520\,Nhits/MeV.

The energy response to the electron energy deposition, the index of refraction, and the absorption length of the LAB-PPO liquid scintillator are investigated in \cite{wan11, wan11_1}. The energy response is linear in the region from 0.4\,MeV to 3.0\,MeV, while below 0.4\,MeV the linearity is lost due to reemission effects and the loss of Cherenkov light (threshold of $\sim$0.2\,MeV).

Finally, the quenching of proton and alpha particles for the unloaded scintillator and the Te-loaded cocktail has been measured in \cite{vK1, vK3}. The nonlinear energy--dependent proton/alpha light output is typically parametrized by Birks' parameter $kB$ \cite{bir64}. Its measurement is extremely important for the development of background rejection techniques as described in Section \ref{sec::Background}. For protons in the unloaded SNO+ scintillator, the value measured in \cite{vK1} is $kB$ = 0.0098 $\pm$ 0.0003\,cm$\cdot$MeV$^{-1}$. The measured value %\cite{vK2} 
for alpha particles is $kB$ = 0.0076 $\pm$ 0.0003\,cm$\cdot$MeV$^{-1}$, corresponding, approximately, to a quenching factor of 10 for energies between 5\,MeV and 9\,MeV. %As a result the light output of an alpha particle of 5\,MeV energy, is the same as that of a beta with 0.5\,MeV kinetic energy (T$_{\beta}$).

\subsection{Process Plant}\label{sec::plant}

The scintillator purification plant of SNO+ is fully described in \cite{ford11, ford15}. It will use the same techniques and has the same cleanliness requirements as the Borexino experiment, by which we expect to reach a purity level of about 10$^{-17}$ g/g$_{LAB}$ for both the $^{238}$U and $^{232}$Th chain \cite{bxo09}, corresponding to 9 counts per day (cpd) for the $^{238}$U chain and 3 cpd for the $^{232}$Th chain. Similar background levels have also been achieved by the KamLAND experiment \cite{Kam03}. A multistage distillation (to remove heavy metals and optical impurities) and a high temperature flash vacuum distillation are initially used to separately purify LAB and PPO. Then the PPO is combined with the LAB, and the scintillator is further purified by a N$_{2}$/steam gas stripping process to remove gases, such as Rn, Ar, Kr, O$_{2}$, and residual water. 

After the detector fill, the entire scintillator volume can be recirculated in about 4 days to enable quasi-batch repurification and \textit{ex-situ} radio-assaying. A rotating-stage liquid-liquid extraction column (water-LAB) and metal scavengers are used to effectively remove metals (K, Pb, Bi, Th, and Ra). Finally, microfiltration is used for removal of suspended fine particles.

During the neutrinoless double-beta decay phase, the tellurium, the water, and the surfactant will be purified prior to addition to the LAB-PPO scintillator. The purification technique for tellurium is described in \cite{yeh15}. It has been designed to remove both the U- and Th-chain impurities and the isotopes produced by cosmogenic neutron and proton spallation reactions while handling and storing tellurium on surface. It consists of a double-pass acid-recrystallization on the earth's surface, for which the overall purification factor reached in U/Th and cosmogenic induced isotopes is $>$10$^{4}$. Since the tellurium purification is expected to happen at the above ground facilities and some isotopes can be cosmogenically replenished even with short time exposures, a second purification stage is needed underground. In this stage telluric acid is dissolved in water at 80\degree C and left to cool to recrystallize without further rinsing. A further purification of about a factor 100 is obtained. Currently, we are investigating the possibility of moving the above ground purification underground, in order to reduce potential recontamination.

The water purification plant at the SNOLAB underground laboratory is based on the SNO light water purification plant, which has been upgraded to improve its performance.

%The purification procedure for the surfactant is still under development. Several tests are on-going in order to select the best purification in terms of background levels and optical properties. 
%The surfactant used to load tellurium is planned to be produced directly at the SNOLAB underground laboratory to avoid any potential neutron and proton induced contamination.
Spike tests have shown that some of the isotopes produced by cosmogenic activation of the surfactant are harder to remove by purification than in the case of telluric acid. The procedure to obtain pure surfactant will therefore be based on its chemical synthesis in a dedicated underground plant.%, based on purified components.

% 1.3$\cdot$10$^{-18}$ g$_{K}$/g$_{LAB}$, which leads to about 23 cpd of $^{40}$K, and less than 100 cpd of $^{85}$Kr and $^{39}$Ar.

\subsection{AV Rope System}
\begin{figure}[t]\centering
\includegraphics[width=6.0 cm]{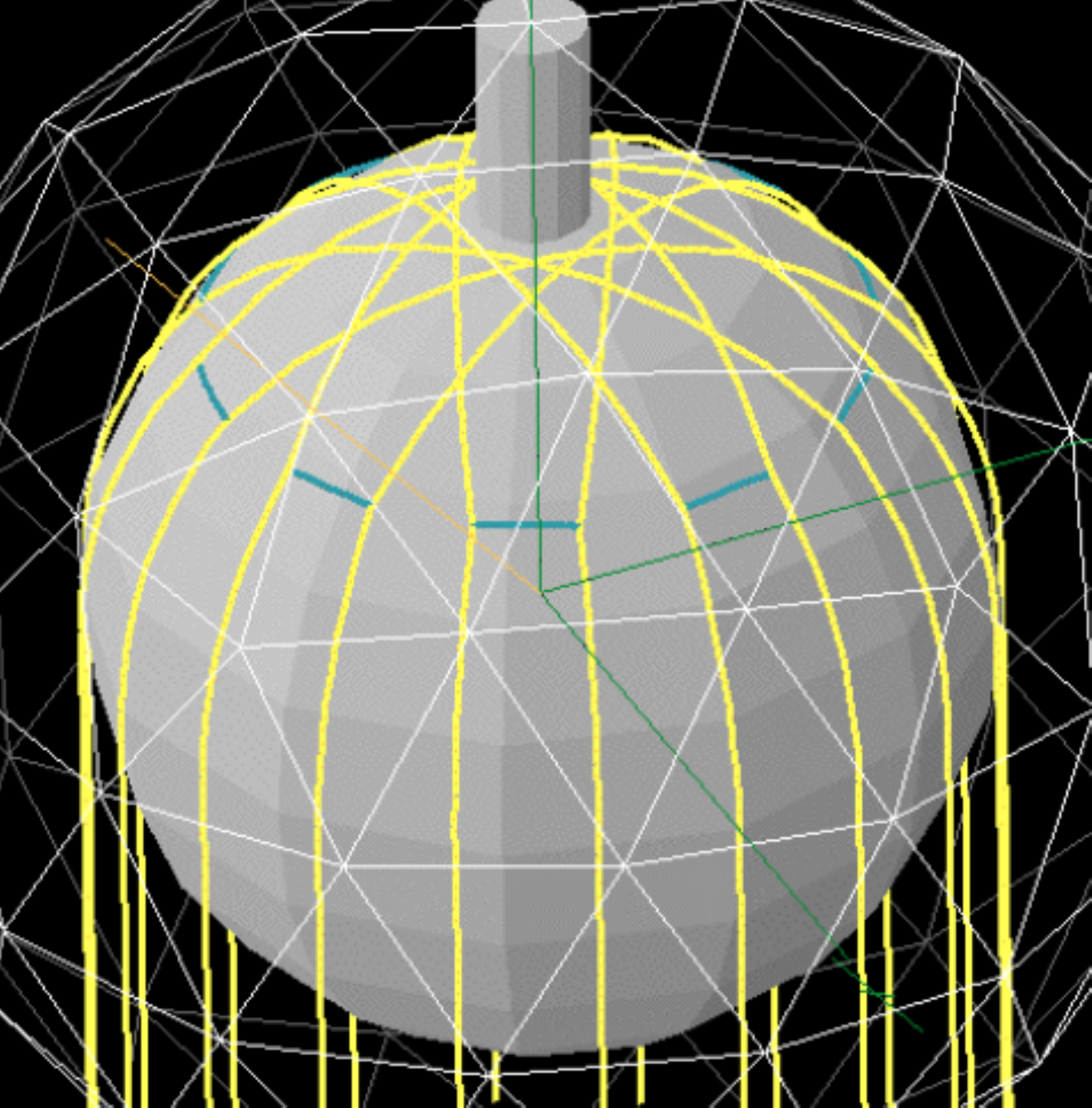} 
\caption{Sketch of the hold down rope system on the top of the acrylic vessel to compensate for the buoyant force that the scintillator produces on the AV.}\label{fig::hold_down}
\end{figure}

The SNO+ liquid scintillator has a lower density ($\rho = 0.86$ g/cm$^{3}$ for LAB-PPO at T = 12\degree C) compared to the surrounding light water, requiring a new hold-down rope system (see Figure\,\ref{fig::hold_down}) to compensate the buoyant force, anchoring the acrylic vessel to the cavity floor. The new hold-down rope system consists of very high purity, high-performance polyethylene fiber (Tensylon) ropes of 38\,mm diameter. %\cite{bialek15}. 
The original hold-up rope system has also been replaced with new Tensylon ropes of 19\,mm diameter in order to reduce the radioactivity contamination.

\begin{table*}[t]
\caption{Calibration sources that are considered for use by the SNO+ experiment. %The label \textit{tag}, means that the beta radiation emitted simultaneously to the gamma radiation is detected via a dedicated scintillator counter embedded in the source container, and it is used as a tag for the decay.
\label{tab::sources}}
\begin{center}
\scalebox{1.0}{
\begin{tabular}{c c c c c c c c}
\hline 
Source & AmBe  & $^{60}$Co & $^{57}$Co & $^{24}$Na & $^{48}$Sc & $^{16}$N & $^{220}$Rn/$^{222}$Rn \\ \hline
Radiation & n, $\gamma$ & $\gamma$ & $\gamma$ & $\gamma$ & $\gamma$ & $\gamma$ & $\alpha, \beta, \gamma$ \\ \hline
Energy [MeV] & 2.2, 4.4 ($\gamma$)  & 2.5 (sum) & 0.122 & 4.1 (sum) & 3.3 (sum) & 6.1 & various \\ \hline
\end{tabular}}
\end{center}
\end{table*}

\subsection{PMTs and Electronics}
SNO+ uses the original 8 inch SNO photomultiplier tubes (Hamamatsu R1408). Each PMT is equipped with a 27\,cm diameter concentrator, increasing the effective photocathode coverage to about 54\%. Faulty PMT bases have been repaired and replaced, and about 9400 PMTs (90 of which are facing outwards) are expected to be in operation at the start of the SNO+ experiment data taking. 

%Differently from the SNO experiment that had a detector threshold of about 3.5\,MeV, 
In SNO+ the use of liquid scintillator as target volume greatly increases the light yield in contrast to the SNO heavy water, allowing the measurement of very low energy signals, like \textit{pp} solar neutrinos (0.4\,MeV end-point energy). Moreover, some of the background event types have high rates of several hundred Hz. For these reasons, the SNO readout boards and the data acquisition system were replaced with new ones capable of a higher bandwidth. New utilities have been added to the SNO+ trigger system which will allow for a more sophisticated use, a flexible calibration interface, and new background cuts to improve the physics sensitivity. The SNO+ trigger window is 400\,ns long, during which time information and charge are collected from every PMT that fired. A dead-time of 30--50\,ns separates two trigger windows \cite{PJ_PhD}. 

In 2012 and 2014, the new electronics and trigger system were tested in runs with the detector empty and nearly half-filled with ultra-pure water (UPW).

\subsection{Cover Gas System}\label{sec::covergas}
As long-lived radon daughters are a potential background for the physics goals of SNO+ (see Section \ref{sec::Background}), the original SNO cover gas system has been upgraded to prevent radon ingress in the detector during operation. It consists of a sealed system filled with high purity nitrogen gas which acts as a physical barrier between the detector and the $\sim$130\,Bq/m$^{3}$ of radon in the laboratory air. A new system of radon tight buffer bags has been designed and installed to accommodate the mine air pressure changes, with the aim of reaching a factor 10$^{5}$ in radon reduction.

\subsection{Calibration Systems}\label{sec::calib}

The SNO+ detector will be calibrated using both optical sources (LEDs and lasers coupled to optical fibers) and radioactive sources (beta, gamma, alpha, and neutron). The optical sources are used to verify the PMT response and to measure \textit{in situ} the optical properties of the detector media, while the radioactive sources are used to check the energy scale, the energy resolution, the linearity of the response, and the detector asymmetries, and to determine the systematic uncertainties and the efficiency of all reconstructed quantities (i.e., energy, position, and direction). Additionally, a system of cameras in underwater enclosures will be used to monitor the position of the acrylic vessel and the hold-down rope system, and to triangulate the positions of the calibration sources inserted into the detector.

The SNO+ calibration hardware has been designed to match the purity requirements of SNO+ and the need to have materials compatible with LAB. The calibration sources will be attached to an umbilical and moved by a system of high purity ropes in order to scan the detector off the central axis in two orthogonal planes.

The set of radioactive sources that are considered for the SNO+ experiment is shown in Table\,\ref{tab::sources}, covering the energy range from 0.1\,MeV to 6\,MeV. In addition, the internal radioactivity can be used to calibrate the detector and check any energy shift or variation of the response during data taking. Typical calibration references are $^{210}$Po-alpha, $^{14}$C-beta, delayed $^{214}$Bi-Po ($^{238}$U chain) and $^{212}$Bi-Po ($^{232}$Th chain) coincidences and muon followers.

The optical calibration hardware consists of internally deployable sources -- a laserball (light diffusing sphere) and a Cherenkov source for absolute efficiency measurements -- and an external system consisting of sets of optical fibers attached to the PSUP in fixed positions, sending pulses from fast LEDs or lasers into the detector. This system allows frequent calibrations of the PMTs response, time, and gain \cite{tellie15}, and measuring the scattering and attenuation length of the scintillator without the need for source insertion.

\subsection{Simulation and Analysis}

A Geant4-based software package RAT (RAT is an Analysis Tool) has been developed to simulate the physics events in the SNO+ detector in great detail, and to perform analyses such as vertex reconstruction.  The RAT simulation includes full photon propagation, from generation via scintillation and Cherenkov processes, through to absorption and detection on the PMTs. The detailed data acquisition and trigger systems are also part of the simulation. Several particle generators have been developed to simulate \obb events, solar neutrinos, geoneutrinos, reactor antineutrinos, supernova neutrinos and antineutrinos. The decay schemes of all relevant background isotopes are also part of the simulation tool. RAT communicates with a database that contains calibration constants and parameters describing the detector status during each run. This includes the optical properties of the various components of the scintillator cocktail, PMT calibration constants, and detector settings such as channel thresholds. Algorithms have been developed to reconstruct event information such as the vertex position, event direction (where relevant), and deposited energy. The SNO+ MC tool is continuously tuned to match newly available measurements.

For all SNO+ physics topics we have run a full Monte Carlo simulation to predict the fraction of background events in the corresponding region of interest (ROI), from which we have evaluated our sensitivities. 

\section{Physics Goals, Current Status, and Run Plan}\label{sec::status}

The primary goal of SNO+ is to search for the neutrinoless double-beta decay of $^{130}$Te. However, it has the potential to explore other physics including the following.
\begin{itemize}
\item \textit{Low Energy pep and CNO Solar Neutrinos} \\
The \textit{pep}-neutrinos can be used to constrain new physics scenarios on how neutrinos couple to matter, while the CNO-neutrino flux can shed light on unresolved questions regarding solar metallicity.
\item \textit{Geoneutrinos} \\
They are produced by the decay of U and Th chains in the Earth's crust and mantle. They can help to understand the heat production mechanisms of the Earth itself. %Flux and distribution of the geo-antineutrinoss depend on the composition of the crust surrounding the detector, which will make SNO+ a perfect candidate for a multi-site measurement along with KamLAND \cite{eno05} and Borexino \cite{bxo13}.
\item \textit{Reactor Antineutrinos} \\
These can be used to better constrain the $\Delta m^{2}_{12}$ neutrino oscillation parameter.%The spectrum depends on the reactor fuel composition, while the oscillation parameters depend on the distance from the source to the detector. The total expected reactor flux in SNO+ is 1/5 of that detected by KamLAND \cite{Kam08}. 
\item \textit{Supernova Neutrinos and Antineutrinos} \\
The ability to detect a galactic supernova provides the potential for improving models of supernova explosions. %SNO+ will be part of the  Supernovae Neutrinos Early Warning System (SNEWS), and has .
\item \textit{Exotic Physics} \\
The low background expected in SNO+ allows searches for processes predicted by physics beyond the standard model (other than 0$\nu\beta\beta$-decay), like invisible nucleon decay, and solar axion or axion-like particle searches. %The detected signal is the emission of a 6.2\,MeV gamma from the de-excitation of the remaining nucleus. SNO+ can set a new limit on the invisible modes of $^{16}$O decays.
\end{itemize}

Currently, the SNO+ cavity is partially filled with ultra-pure water. %with the corresponding level of water in the acrylic vessel approaching one-third. 
The upgrades to the SNO+ detector are nearly completed with a few items to be finished before the start of data taking. The detector parts that need to be finalized are the installation of the calibration system, underwater cameras, and the calibration optical fibers in most of the positions above the SNO+ equator, and the replacement of the PMTs. The installation will proceed along with the rise of the water level in the cavity. The scintillator plant is nearly completed. The newly installed electronic and trigger system and part of the optical calibration system have been tested in air and with the partially water-filled detector.\\

The data taking period of SNO+ will be divided into three main phases:
\begin{description}
\item[Water phase:] In this phase, the acrylic vessel will be filled with about 905\,tonnes of ultra-pure water and data taking will last for a few months. The main physics goals will be a search for exotic physics, including solar axion-like particles and invisible nucleon decay in $^{16}$O, the watch for supernova neutrinos, and the detection (potentially) of reactor antineutrinos. During this phase, the detector performance, the PMT response and the data acquisition system characteristics will be tested. Optical calibrations to test the response of the PMT concentrators and the attenuation of the external water and the acrylic will be performed. The backgrounds coming from external sources, like external water, PMT array, hold-down ropes, and the acrylic vessel, will be characterized. 
\item[Pure scintillator phase:] In this phase, the detector will be filled with about 780 tonnes of LAB-PPO liquid scintillator and data taking will last for a few months. The physics topics covered are the measurement of the low energy solar neutrinos,  the measurement of geo and reactor antineutrinos, and the supernova neutrino watch. This phase will also be used to verify the optical model and the detector response and to characterize the backgrounds due to internal and external radioactive sources. 
\item[Te-loading phase:] This phase is foreseen to start in 2017 and last for about 5 years. In this phase, also called Phase I, about 2.3 tonnes of natural tellurium (0.3\% loading by weight) will be added to the detector for the search for the \obb of $^{130}$Te. Simultaneously, geo and reactor neutrinos can be observed, and the detector will be live to a potential supernova.
\end{description}

The physics program and capabilities of SNO+ will be discussed in Sections \ref{sec::doublebeta} to \ref{sec::exotic}.

\section{Backgrounds}\label{sec::Background}

The background sources of the SNO+ experiment can be divided into two main categories: internal and external. Internal backgrounds are all the non-signal interactions that occur inside the AV (R $<$ 6\,m). External backgrounds are the interactions that are produced in the region outside the target volume but that can propagate or are reconstructed within it. Full Monte Carlo simulations, along with \textit{ex-situ} assays are used to explore the different background sources and develop rejection techniques.

In the following subsections the various background sources are presented: internal $^{238}$U chain (Section \ref{sec::238U}), $^{210}$Bi and $^{210}$Po decays (Section \ref{leaching}), internal $^{232}$Th chain (Section \ref{sec::232Th}), internal $^{40}$K, $^{39}$Ar, and $^{85}$Kr decays (Section \ref{sec::40K}), cosmogenically induced isotopes (Section \ref{sec::cosmo}), ($\alpha$,n) reactions (Section \ref{sec::an}), pile-up events (Section \ref{sec::pileup}), and external backgrounds (Section \ref{sec::external}).

\subsection{Internal $^{238}$U Chain}\label{sec::238U}

\begin{figure}[t]
\centering
\includegraphics[scale=0.5]{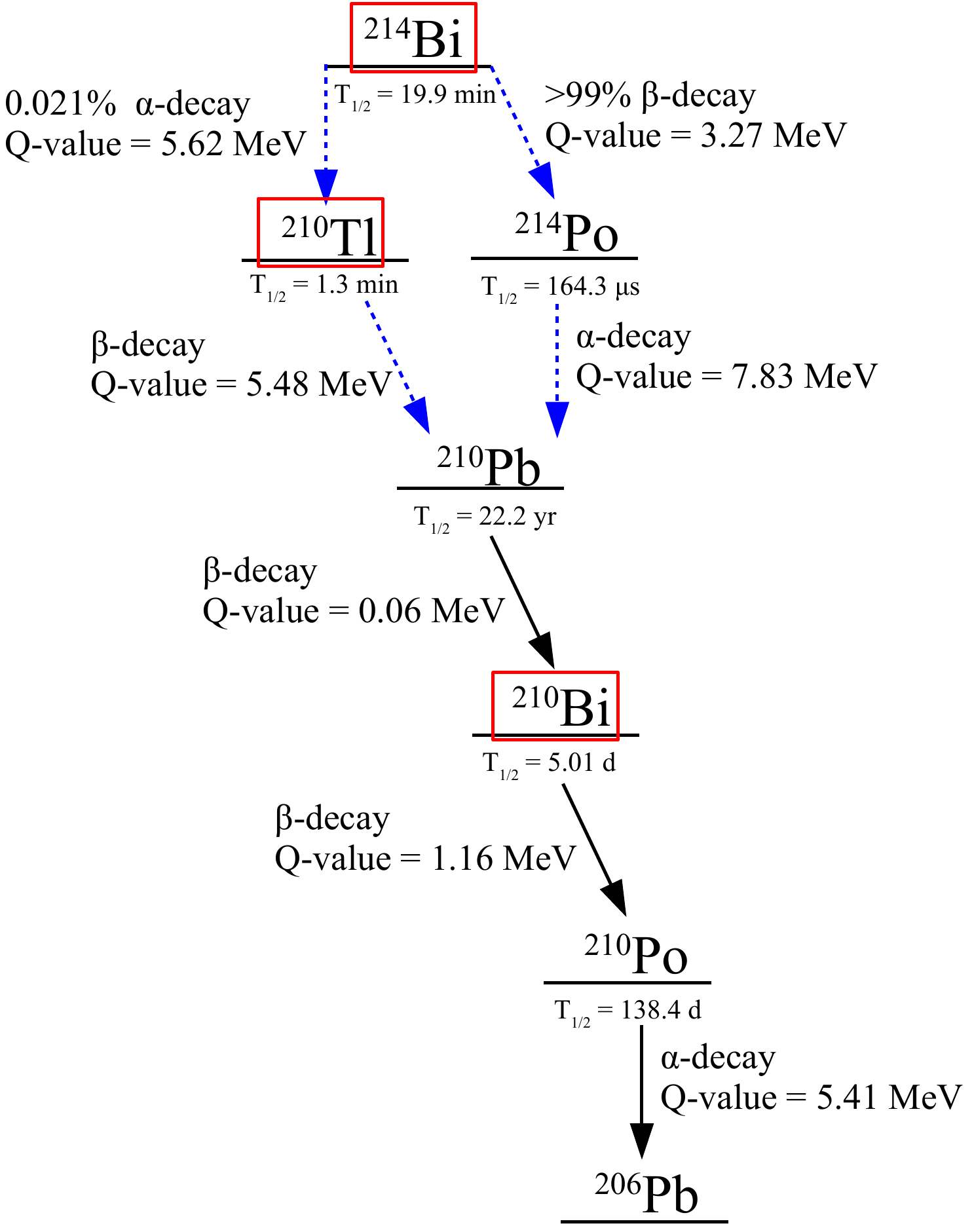} 
\caption{Part of $^{238}$U-decay chain relevant for SNO+ with Q-values (total kinetic energy released in the ground state - ground state transition), half-life and decay modes \cite{nndc}. The red squares highlight the nuclides of most concern: $^{214}$Bi, $^{210}$Tl, and $^{210}$Bi. The decays used for $\alpha$-$\beta$ and $\beta$-$\alpha$ coincidence techniques are shown with a blue arrow (dash-dotted line). \label{fig::238U}}
\end{figure}

$^{238}$U (T$_{1/2}$ = 4.47$\times$10$^{9}$\,yr) is a naturally occurring radioisotope present in the liquid scintillator. The part of the decay chain relevant for SNO+ is shown in Figure\,\ref{fig::238U}. The $^{238}$U daughters of most concern are $^{214}$Bi, $^{210}$Tl, and $^{210}$Bi (see Section \ref{leaching}). Secular equilibrium with the top part of the chain is assumed through the paper unless otherwise noted.

$^{214}$Bi (T$_{1/2}$ = 19.9\,min) beta-decays to $^{214}$Po with a Q-value of 3.27\,MeV in 99.979\% of the cases. This decay can be tagged using the $^{214}$Po alpha-decay (T$_{1/2}$ = 164.3\,$\mu$s, E$_{\alpha}$ = 7.7\,MeV), during both the pure scintillator and the Te-loaded phase. In the pure scintillator phase, the $\beta$--$\alpha$ delayed coincidence will be used to measure the concentration of the $^{238}$U-chain contaminants. $^{214}$Bi is expected to be in secular equilibrium with the top part of the $^{238}$U chain for most of the data taking period. This equilibrium can be broken by radon ingress into the detector during calibration campaigns, or from emanation by the calibration hardware materials. However, for non-continuum sources of radon, due to the short half-life of $^{214}$Bi, equilibrium will be restored in a few weeks' time. In SNO+ the presence of the cover gas on the top of the detector provides an efficient barrier against laboratory air, highly reducing the radon ingress into the detector (see Section \ref{sec::covergas}). Additionally, most of the radon short-lived daughters decay in the cover gas region or in the detector neck; thus they do not reach the fiducial volume. 

During the Te phase, the delayed coincidence technique will be used to reject $^{214}$Bi events that fall into the region of interest (ROI) for the \obb search.

Usually, Bi-$\beta$ and Po-$\alpha$ are separated by more than 250\,ns and the SNO+ detector records them as two separate events. The secondary events (alpha candidates) are identified by applying an energy cut around the alpha energy, shifted due to quenching to $\sim$0.8\,MeV electron equivalent energy, and by the short time separation from the preceding event. To reduce the misidentification of the events due to other decays occurring in the same energy region during the coincidence window, a position cut can also be applied. An $\alpha$-$\beta$ classification algorithm has been developed to further reduce the misidentification by classifying the events as $\alpha$-like or $\beta$-like based on the hit-time distribution.

Occasionally, the beta and the alpha decays are separated by less than 250\,ns and they may be recorded as a single event by the SNO+ detector. These events are important for the \obb phase as they may fall into the ROI. In this case, the rejection technique is based on the distortion in the time distribution of the light detected by the PMTs compared to the case of a single interaction. This rejection technique is enhanced if a pulse shape analysis can be applied to distinguish beta from alpha events.

In 0.021\% of the cases $^{214}$Bi alpha-decays to $^{210}$Tl (T$_{1/2}$ = 1.3\,min), which beta-decays to $^{210}$Pb with a Q-value of 5.5\,MeV. Due to the small branching ratio this route is less important than the previous one. An $\alpha$--$\beta$ delayed coincidence, similar to the $\beta$--$\alpha$ one, can be applied. However, due to the longer half-life of $^{210}$Tl, the mis-tagging probability is larger with respect to the $^{214}$Bi-Po one which may result in a larger signal sacrifice.\\

Based on Borexino Phase-I achievements \cite{bxo09}, the purity level aimed (target level) in the LAB-PPO scintillator for the $^{238}$U chain is 1.6$\times$10$^{-17}$ g/g (see Table\,\ref{tab::back_sources_int}). During the Te-loaded phase, the addition of the isotope, the water, and the surfactant to LAB will worsen the mixture purity, but we will maintain a strict target level of 2.5$\times$10$^{-15}$ g/g (see Table\,\ref{tab::back_sources_int}).

\begin{table}[t]
\caption{Target levels, in g/g, and corresponding decay rates for the internal $^{238}$U- and $^{232}$Th-chain contaminants in the various SNO+ phases. Secular equilibrium has been assumed for all the isotopes except $^{210}$Pb, $^{210}$Bi, and $^{210}$Po. The levels of $^{210}$Bi and $^{210}$Po during the pure scintillator phase and the Te-loaded phase are expected to be out of secular equilibrium due to the intrinsic scintillator contamination and the leaching off of the AV surface. For the 0.3\% Te-loaded scintillator the tellurium/polonium affinity component is also included in the $^{210}$Po decays/yr (see text).\label{tab::back_sources_int}}
\begin{minipage}{8.5cm}
\begin{center}
\scalebox{1.0}{
\begin{tabular}{c c c}
\hline 
Source & Target [g/g] & Decays/yr \\ \hline \hline
\multicolumn{3}{|c|}{\textit{Internal H$_{2}$O, water phase}} \\ \hline
$^{238}$U chain  & 3.5$\times$10$^{-14}$ & 1.2$\times$10$^{7}$ \\ 
$^{232}$Th chain & 3.5$\times$10$^{-15}$ & 4.1$\times$10$^{5}$\\  \hline
\multicolumn{3}{|c|}{\textit{LAB-PPO, pure scintillator phase}} \\ \hline
$^{238}$U chain & 1.6$\times$10$^{-17}$ & 4900 \\ 
$^{232}$Th chain & 6.8$\times$10$^{-18}$ & 700\\ 
$^{210}$Bi  & - & 7.6$\times$10$^{8}$ \footnote{\label{note1}Expected number of events in the first year after 9 months of water phase.}\\ 
$^{210}$Po  & - & 7.8$\times$10$^{8}$ \footnoteref{note1}\\ \hline
\multicolumn{3}{|c|}{\textit{0.3\% Te-loaded scintillator, Te phase}} \\ \hline
$^{238}$U chain & 2.5$\times$10$^{-15}$ & 7.6$\times$10$^{5}$ \\ 
$^{232}$Th chain & 2.8$\times$10$^{-16}$ & 2.8$\times$10$^{4}$\\ 
$^{210}$Bi  & - & 7.9$\times$10$^{9}$ \footnote{\label{note2}Expected number of events in the first year after 9 months of water phase followed by 6 months of pure scintillator phase.}\\ 
$^{210}$Po  & - & 9.5$\times$10$^{9}$ \footnoteref{note2}\\ \hline
\end{tabular}}
\end{center}
\end{minipage}
\end{table}

\subsection{$^{210}$Bi and $^{210}$Po Backgrounds}\label{leaching}

The ingress of $^{222}$Rn into the SNO+ detector can break the secular-equilibrium in the $^{238}$U chain at $^{210}$Pb (T$_{1/2}$ = 22.2 yr, Q-value = 0.06\,MeV), resulting in a higher concentration of this isotope. Even if $^{210}$Pb is not a direct background for the SNO+ experiment, its daughters $^{210}$Bi (T$_{1/2}$ = 5.0\,d, Q-value = 1.16\,MeV) and $^{210}$Po (T$_{1/2}$ = 138.4\,d, E$_{\alpha}$ = 5.3\,MeV, shifted to $\sim$0.5\,MeV electron equivalent energy) are potentially relevant for the various physics searches. $^{210}$Bi-beta decays are the main background for the CNO-$\nu$ measurement, as they have similar spectral shapes, while the $^{210}$Po-alpha decay  is a background for the $\beta$--$\alpha$ and $\alpha$--$\beta$ delayed coincidences, resulting in mis-tagging and potential signal sacrifice. Additionally, the emitted alphas can interact with the atoms in the scintillator producing neutrons as described in Section \ref{sec::an}. The cover gas system placed at the top of the acrylic vessel greatly reduces the radon ingress into the detector. Furthermore, the majority of short-lived daughters decay before reaching the fiducial volume. However, due to its long half-life, $^{210}$Pb is not attenuated by the presence of the detector neck and reaches the target volume.

%%As described in Section \ref{sec::Background} the cover gas system at the top of the SNO+ detector highly reduces the ingress of Rn (factor 10$^{4}$--10$^{5}$). Short-lived Rn daughter like $^{214}$Bi and $^{210}$Tl are further reduced by the negligible convection in the vessel chimney and generally decay before reaching the fiducial volume. However, due to the long half life a constant flux of $^{210}$Pb may reach the fiducial volume. 
$^{210}$Pb and its daughters may also leach from materials that are in contact with the liquid scintillator.
Radon daughters deposited on the material's surface can implant by alpha recoil to a depth of a few hundred nm, where they eventually decay to $^{210}$Pb. $^{210}$Pb, $^{210}$Bi, and $^{210}$Po atoms might then leach off when the liquid scintillator mixture is in contact with the surface. This process can happen, for instance, during the handling and storing of the liquid scintillator, resulting in rates of $^{210}$Pb, $^{210}$Bi, and $^{210}$Po out of equilibrium with the $^{238}$U chain. Concentrations of $^{210}$Bi and $^{210}$Po different from each other and the rest of the $^{238}$U chain have been seen by the Borexino experiment \cite{bxo091}. The levels initially measured by Borexino for these two isotopes are included in Table\,\ref{tab::back_sources_int}.

An additional source of $^{210}$Pb, $^{210}$Bi, and $^{210}$Po is leaching from the internal surface of the AV, where radon daughters have implanted during the construction of SNO and when the detector was empty after draining the heavy water. This may create a continuous source of $^{210}$Pb, $^{210}$Bi, and $^{210}$Po during the data taking period for all SNO+ phases. Leaching rates depend on several factors, like temperature, implantation depth, type of liquid in contact with the surface, and initial surface activity. The leaching rate of $^{210}$Pb and its daughters for all the scintillator mixtures and the ultra pure water at different temperatures have been measured in bench top tests. With a measured activity of about 1\,kBq on the inner AV surface, the activity of $^{210}$Pb daughters leached in the scintillator media might be as high as a few hundred Bq depending on the duration of the data taking period. The activity of the backgrounds leached in the scintillator is expected to increase with time, while that of inner surface events is expected to decrease.

In the Te-loaded phase, an additional source of $^{210}$Po is the tellurium itself. The CUORE collaboration has shown \cite{cuore} that due to the chemical affinity between tellurium and polonium this element may still be present in tellurium after the crystal production process. In our background estimations we assume an additional $^{210}$Po activity of 0.06\,Bq/kg$_{Te}$, based on CUORE measurements. These decays, however, are not supported by $^{210}$Pb and are considerably reduced, to about 16\% of the initial activity, in a year after tellurium production. This contribution is included in the purity levels of $^{210}$Po shown in Table\,\ref{tab::back_sources_int}.

\subsection{Internal $^{232}$Th Chain}\label{sec::232Th}

\begin{figure}[t]
\centering
\includegraphics[scale=0.5]{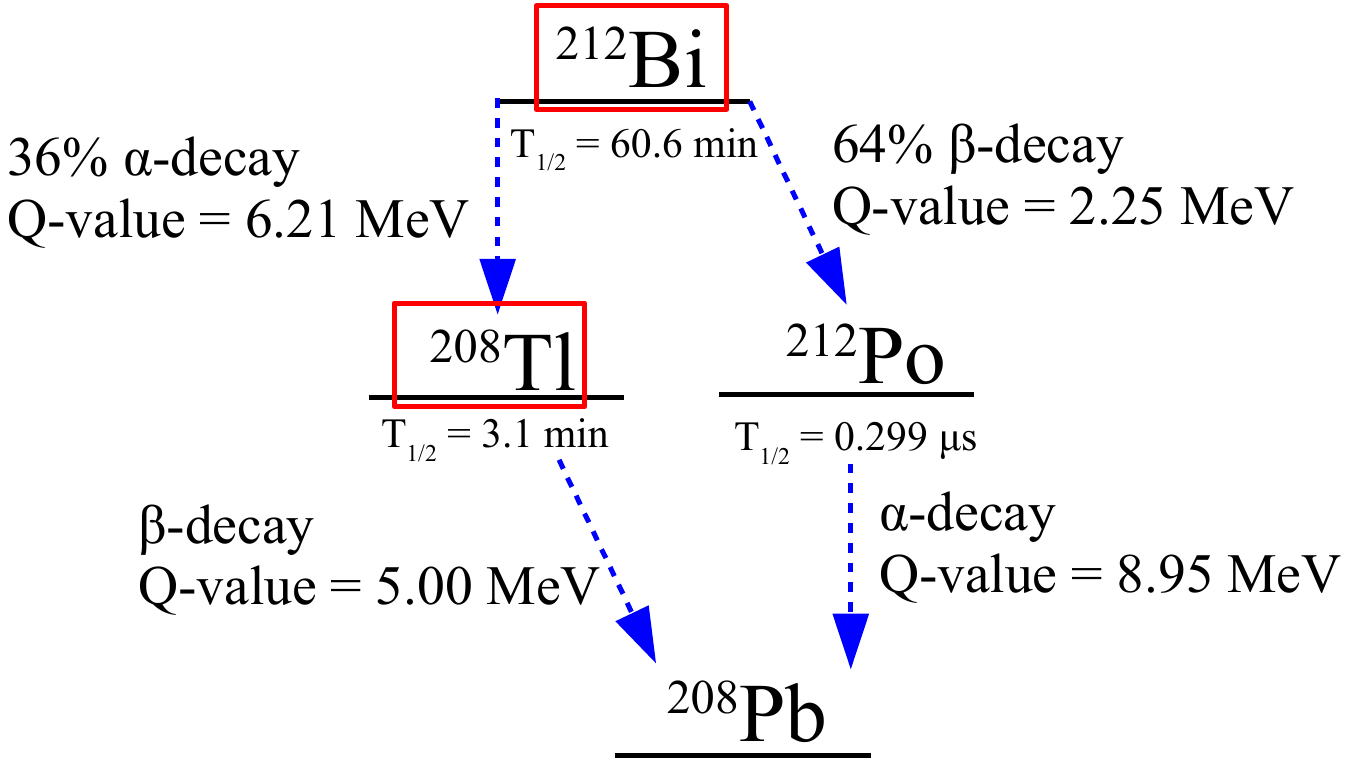} 
\caption{Part of $^{232}$Th-decay chain relevant for SNO+ with Q-values, half-life and decay modes \cite{nndc}. The red squares highlight the most important nuclides: $^{212}$Bi and $^{208}$Tl. The decays used for $\alpha$-$\beta$ and $\beta$-$\alpha$ coincidence techniques are shown with a blue arrow (dash-dotted line). \label{fig::232Th}}
\end{figure}

$^{232}$Th (T$_{1/2}$ = 1.4$\times$10$^{10}$ yr) is also a naturally occurring radioisotope present in the liquid scintillator. The daughters of most concern are $^{212}$Bi and $^{208}$Tl (see Figure\,\ref{fig::232Th}). 

$^{212}$Bi (T$_{1/2}$ = 60.6\,min) beta-decays to $^{212}$Po (T$_{1/2}$ = 300\,ns) with a Q-value of 2.25\,MeV in 64\% of the cases. As for the $^{214}$Bi$\rightarrow^{214}$Po decay, many events can be selected using a $\beta$--$\alpha$ delayed coincidence, which is used to extract the concentration of the $^{232}$Th-chain contaminants in equilibrium in the pure scintillator. Nearly 45\% of the $^{212}$Bi$\rightarrow^{212}$Po decays fall in the same trigger window and are a potential background for the \obb search. These can be rejected using the PMT timing distribution.

In the remaining 36\% of the cases $^{212}$Bi alpha-decays to $^{208}$Tl (T$_{1/2}$ = 3.0\,min), which beta-decays to $^{208}$Pb with a Q-value of 5.0\,MeV. An $\alpha$--$\beta$ delayed coincidence can be applied to identify the $^{208}$Tl events as for the $^{210}$Tl case.\\

The LAB-PPO scintillator target level for the $^{232}$Th chain is 6.8$\times$10$^{-18}$ g/g (based on \cite{bxo09}), while the target level for the Te-loaded scintillator is 2.8$\times$10$^{-16}$ g/g (see Table\,\ref{tab::back_sources_int}).

\subsection{Internal $^{40}$K, $^{39}$Ar, and $^{85}$Kr Backgrounds}\label{sec::40K}

Other internal backgrounds are important for solar neutrino and other measurements.

$^{40}$K (T$_{1/2}$ = 1.248$\times 10^{9}$ yr) has a very distinctive energy spectrum, having both a beta component and a gamma peak at 1.46\,MeV. Due to the long half-life, it is naturally present in the scintillator and detector materials.

$^{39}$Ar (T$_{1/2}$ = 269 yr), and $^{85}$Kr (T$_{1/2}$ = 10.8 yr) decay with a Q-value of 0.565\,MeV and of 0.687\,MeV, respectively. The amount of these isotopes can be reduced by minimising the contact time of LAB with air and thoroughly degassing the scintillator. %A tagging technique for $^{85}$Kr, based on the identification of the emitted gamma (BR=0.4\%, T$_{1/2}$=1.02\,$\mu$s), is under development. %The current efficiency is around 20\%.

\begin{table*}[t]
\caption{$^{238}$U- and $^{232}$Th-chain levels for external background sources. Shown are measured levels and expected decay rates. \label{tab::back_sources_ext}}
\begin{minipage}{18cm}
\begin{center}
\scalebox{1.0}{
\begin{tabular}{c c c}
\hline 
Source & Measured levels & Decays/yr \\ \hline
Internal ropes &  $^{214}$Bi: $(2.8\pm5.4)\times10^{-10}$g$_{U}$/g \cite{Gesnolab} & $4966$  \\
 & $^{208}$Tl: $<2.0\times10^{-10}$g$_{Th}$/g \cite{Gesnolab} & $< 418$ \\
Hold-down ropes & $^{214}$Bi: $(4.7\pm3.2)\times10^{-11}$g$_{U}$/g \cite{Gesnolab} & $4.06\times10^{6}$  \\
 & $^{208}$Tl: $(2.27\pm1.13)\times10^{-10}$g$_{Th}$/g\cite{Gesnolab} & $2.32\times10^{6}$ \\
Hold-up ropes & $^{214}$Bi: $(4.7\pm3.2)\times10^{-11}$g$_{U}$/g  \cite{Gesnolab} & $8.34\times10^{5}$  \\
 & $^{208}$Tl: $(2.27\pm1.13)\times10^{-10}$g$_{Th}$/g \cite{Gesnolab} & $4.78\times10^{5}$ \\
Water Shielding &$^{214}$Bi: $2.1\times10^{-13}$g$_{U}$/g \cite{sno05}& $1.32\times10^{8}$ \\
& $^{208}$Tl: $5.2\times10^{-14}$g$_{Th}$/g \cite{sno05} & $3.92\times10^{6}$ \\
Acrylic Vessel  & $^{214}$Bi: $<1.1\times10^{-12}$g$_{U}$/g \footnote{\label{note3}Assumed $1.0\times10^{-12}$g/g}\cite{sno00} & $1.28\times10^{7}$ \\
& $^{208}$Tl: $<1.1\times10^{-12}$g$_{Th}$/g \footnoteref{note3}\cite{sno00} & $1.50\times10^{6}$ \\
Acrylic Vessel External Dust\footnote{It is assumed that the top hemisphere of the external AV surface is not cleaned, while the bottom hemisphere is at target level}  & $^{214}$Bi: $(1.1\pm0.1)\times10^{-6}$g$_{U}$/g \cite{norite} & $7.8\times10^{5}$  \\%$1.1\times10^{-6}$g/g \cite{nasimbg} & $2.0\times10^{5}$ \\
& $^{208}$Tl: $(5.6\pm0.5)\times10^{-6}$g$_{Th}$/g \cite{norite} & $4.6\times10^{5}$  \\%$5.6\times10^{-6}$g/g \cite{dustbg} & $1.2\times10^{5}$ \\
%& & $^{210}$Po & $1.9$ Bq/m$^{2}$  & $1.2\times10^{5}$ \\
Acrylic Vessel Internal Dust& $^{214}$Bi: $(1.1\pm0.1)\times10^{-6}$g$_{U}$/g \cite{norite} & $4.15\times10^{4}$  \\
& $^{208}$Tl: $(5.6\pm0.5)\times10^{-6}$g$_{Th}$/g \cite{norite} & $2.48\times10^{4}$ \\
PMTs & $^{214}$Bi: 100$\times10^{-6} g_{U}/$PMT \cite{sno00} & $3.7\times10^{11}$ \\
& $^{208}$Tl: 100$\times10^{-6} g_{Th}/$PMT \cite{sno00} & $4.4\times10^{10}$  \\  \hline
\end{tabular}}
\end{center}
\end{minipage}
\end{table*}

\subsection{Cosmogenically Induced Backgrounds}\label{sec::cosmo}

Besides the natural radioactivity present in the scintillator, LAB can be activated by cosmic ray neutrons and protons while it is above ground. The main expected background is $^{7}$Be (T$_{1/2}$ = 53.2 d, EC-decay with a 0.48\,MeV gamma), with a maximum production rate at sea level (neutron and proton flux from \cite{arm, geh}) of about 1 kHz for 780\,t of liquid scintillator. More than 99\% of the produced $^{7}$Be can be efficiently removed by the scintillator purification plant. %In the Te-loaded cocktail, about 30\,Bq of $^{7}$Be may result from the cosmogenic activation of the surfactant while it is transported at site. Currently, we don't expect to remove this activity, however, both the rate and the energy are not expected to be a problem for the \obb search. In addition, the 0.48\,MeV gamma can be used as an internal calibration source in the first months of data taking.

$^{14}$C (T$_{1/2}$ = 5700\,yr, Q-value = 0.16\,MeV) is naturally present in the liquid scintillator. It is a direct background for the very low energy \textit{pp} neutrino measurements and may contribute to pile-up backgrounds (see Section \ref{sec::pileup}). In SNO+, we expect a $^{14}$C/$^{12}$C ratio of the order of 10$^{-18}$, similar to what was observed in the Borexino test facility \cite{bor98}, corresponding to a decay rate of a few hundred Hz. This is a reasonable assumption as in both cases the liquid scintillator is obtained from old oil fields, in which most of the $^{14}$C has decayed away. The amount of $^{14}$C produced by cosmogenic activation of LAB during transport to site is negligible in comparison.

$^{11}$C (T$_{1/2}$ = 20\,min, Q-value = 1.98\,MeV) is mainly produced by muon interactions with the carbon nuclei of the liquid scintillator. We expect a total of (1.14\,$\pm$\,0.21)$\times 10^{3}$\,decays/kt/yr during operation, extrapolated from KamLAND data in \cite{kam10}. This is about a factor 100 less than what was observed in Borexino \cite{bxo11} due to the deeper underground location. A threefold coincidence tagging technique, like the one developed by Borexino \cite{gal05}, together with an electron-positron discrimination analysis \cite{gal11}, will further reduce these events.

Other muon induced backgrounds are generally very short lived (milliseconds to seconds half-life) and can be rejected by vetoing the detector for a few minutes after each muon event.

Important cosmogenic-induced backgrounds are isotopes produced by spallation reactions on tellurium while it is stored on surface \cite{lozza}, like $^{124}$Sb (T$_{1/2}$ = 60.2\,d, Q-value = 2.90\,MeV), $^{22}$Na (T$_{1/2}$ = 950.6\,d, Q-value = 2.84\,MeV), $^{60}$Co (T$_{1/2}$ = 1925\,d, Q-value = 2.82\,MeV), $^{110m}$Ag (T$_{1/2}$ = 249.8\,d, Q-value = 2.89\,MeV, E$_{parent}$(level) = 0.118\,MeV) and $^{88}$Y (T$_{1/2}$ = 106.6\,d, Q-value = 3.62\,MeV).
We have developed a purification technique \cite{yeh15} (see Section \ref{sec::plant}) that, together with underground storage, reduces the cosmogenic-induced background on tellurium to a negligible level.

%$^{22}$Na may also be produced by cosmogenic activation of the surfactant used to load tellurium, during transport to site. We are currently investigating various strategies to limit this background, which may include the production of the surfactant underground.
%Once underground the production rate on tellurium is mainly due to muon induced neutrons and leads to less than 1 event/yr in the full SNO+ detector (full energy range).

%The sulphonate-based surfactant can also be activated by cosmogenic neutrons and protons with the production of $^{22}$Na during transportation at site. Currently a purification technique to be used underground to target this background is under development.

\subsection{($\alpha$,n) Backgrounds}\label{sec::an}

Neutrons can be produced in the liquid scintillator by ($\alpha$,n) reactions on $^{13}$C or $^{18}$O atoms, muon interactions in the scintillator volume, $^{238}$U fission, and ($\gamma$,n) reactions for E$_{\gamma}>$ 3\,MeV. Excluding the muon induced neutrons, the most prominent neutron source inside the scintillator volume is the $\alpha$+$^{13}$C\,$\rightarrow^{16}$\,O+$n$ reaction (E$_{thr.}$ = 0.0\,keV), which is a potential background for both the \obb search and the antineutrino measurement. The main source of alpha particles in the various scintillator mixtures is $^{210}$Po. Other U- and Th- chain's alpha emitters form a negligible contribution, as they are expected to be $\sim$4 orders of magnitude less abundant.

Neutrons produced in ($\alpha$,n) reactions will scatter from protons during the thermalization process, resulting in recoils emitting scintillation light. The visible proton energy together with the energy lost by the alphas before interaction is the prompt signal. If the isotope is in an excited state, the emitted deexcitation gammas are also part of the prompt signal. The thermalized neutrons in $>$99\% of the cases are eventually captured by hydrogen atoms with the emission of the characteristic 2.22~MeV $\gamma$. In the remaining $\sim$1\% of the cases the thermal neutron is captured either on tellurium isotopes, producing mainly a 0.6\,MeV gamma, or on $^{12}$C, producing a 4.95\,MeV gamma. The prompt and the delayed signal can be used to reject the ($\alpha$,n) background using a delayed coincidence technique similar to that of $\beta$-$\alpha$ events.

\subsection{Pile-Up Backgrounds}\label{sec::pileup}

A pile-up event occurs when two or more decays (signal or background or a mixture) happen in the same trigger window and thus are potentially detected as a single event with energy equal to the sum of the single energies. Pile-up events become important when the event rate of one or all of the contributing decays is very high (hundreds of Hz), like $^{14}$C decays or $^{210}$Bi or $^{210}$Po. A rejection technique, using the distortion of the timing, is used to efficiently reduce these backgrounds \cite{Aru14, Aru14-2}.

\subsection{External Backgrounds}\label{sec::external}

Sources of external background include the hold-down and hold-up ropes, the PMT array, the AV bulk, and the external water (see Table\,\ref{tab::back_sources_ext}). Radioactive decays occur outside the scintillator volume, so the main concerns for the signal extraction analysis are the high energy gammas and betas emitted by $^{214}$Bi, $^{208}$Tl, and $^{40}$K decays. External background events reconstructing inside the AV can be greatly reduced by applying a fiducial volume cut. Events can be further reduced using the PMT time distribution. \textit{In-situ} analysis during the water phase and the pure liquid scintillator phase will help to constrain the external backgrounds for the Te-loaded phase.

\section{$^{130}$Te Neutrinoless Double-Beta Decay}\label{sec::doublebeta}

The main goal of the SNO+ experiment is the search for neutrinoless double-beta decay of $^{130}$Te (Q-value = 2527.518 $\pm$ 0.013\,keV \cite{red09}) by loading large quantities of the isotope into the liquid scintillator volume. This approach has several advantages: (1) external backgrounds can be removed by fiducialization, (2) internal and external background levels can be measured before and after the isotope deployment, allowing identification and removal of possible contamination, (3) internal backgrounds can be tagged by coincidences or particle identification, (4) the detector response can be tested with and without the isotope, (5) the spatial distribution of most background isotopes in a liquid is known to be uniform, (6) the loading can be easily and affordably scaled up or (7) changed to another isotope, and (8) tellurium and scintillator can be removed and repurified if high levels of backgrounds are found.

\begin{figure}[t]
\centering
\includegraphics[scale=0.35]{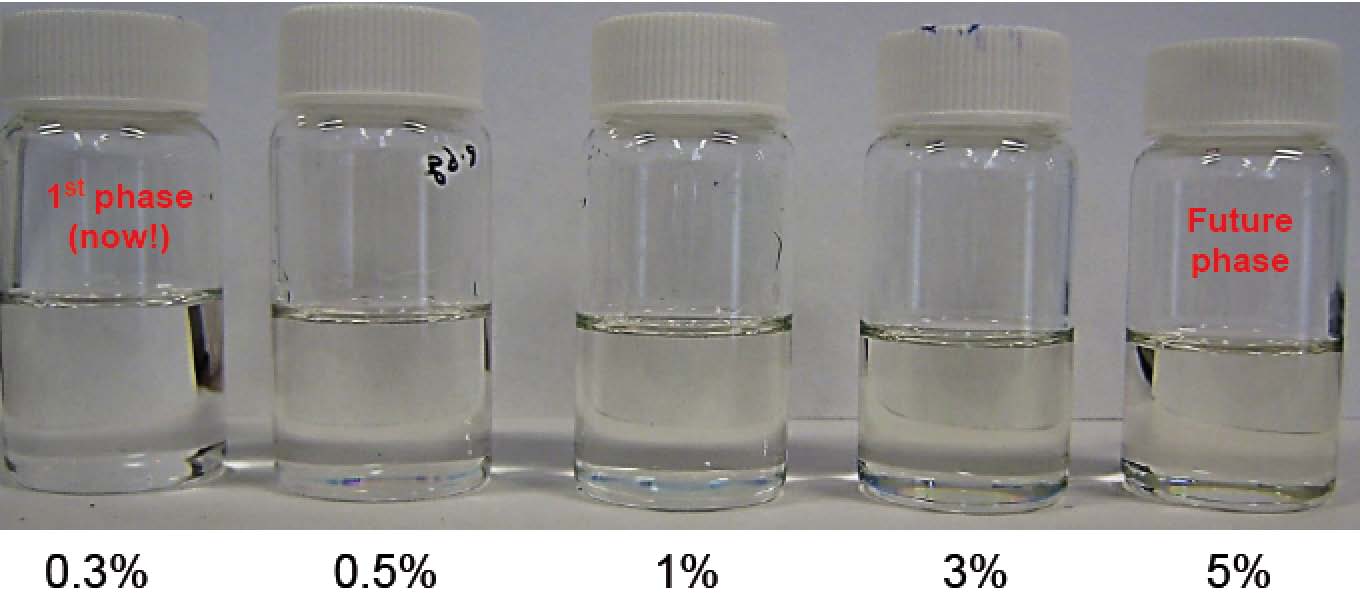} 
\caption{TeLS samples from the investigation of higher tellurium loading in LAB scintillator. The samples increase in loading from 0.3\% (by weight) on the left to 5\% on the right.\label{fig::loading}}
\end{figure}

The choice of $^{130}$Te as the preferred $0\nu\beta\beta$ candidate is the result of an extensive investigation by the SNO+ collaboration. The decision was based on several factors, including the following points:

\begin{enumerate}
\item $^{130}$Te has a large natural abundance of 34.08\%, which allows loading of several tonnes of isotope without enrichment.
\item The measured half-life of the $^{130}$Te $2\nu\beta\beta$ decay is ($7.0\pm0.9\mathrm{(stat)}\pm1.1\mathrm{(syst)})\times10^{20}$\,yr \cite{nemo3}, one of the longest of all the $0\nu\beta\beta$ isotopes. This is particularly important for liquid scintillator-based experiments, as the energy resolution is usually some hundreds of keV.
\item An innovative loading technique has been developed, which enables deployment of up to 5\% (by weight) of natural tellurium while maintaining good light transmission, minimal scattering, and an acceptable light yield (see Section \ref{sec::teloading}). The 0.3\% tellurium scintillator cocktail (TeLS) has been proven to be stable for a period of over two years. In Figure\,\ref{fig::loading}, various SNO+ loaded cocktails are shown. Cocktails with higher loading still maintain good optical transparency.
\item The TeLS does not present inherent optical absorption lines in the visible wavelength range, such that a secondary wavelength shifter may be added to the cocktail to better match the SNO+ PMT response.
\end{enumerate}

%With the initial natural Tellurium concentration of 0.3\% (by weight) in Phase I, corresponding to about 800 kg of $^{130}$Te, SNO+ will reach a sensitivity to the $0\nu\beta\beta$ half life of the order of $10^{26}$ years in 5 live-years. Sensitivity in a subsequent second phase (Phase II) with loading ten times higher is expected to cover the majority of the inverted hierarchy region \cite{whitepaper}.

\subsection{Backgrounds}

For the $^{130}$Te $0\nu\beta\beta$ search, an asymmetric region of interest (ROI) is defined, which extends from $-0.5\sigma$ to $1.5\sigma$ around the Gaussian signal peak. For the 0.3\% Te-loaded cocktail with a light yield of 200\,Nhits/MeV (see Section \ref{sec::teloading}) the energy resolution at 2.5\,MeV is $\sim$270\,keV (FWHM), while the averaged position resolution at the same energy is $\sim$15\,cm at the detector's center. An asymmetric ROI retains most of the 0$\nu\beta\beta$ decays but considerably reduces the backgrounds from $2\nu\beta\beta$ and low energy $^{238}$U- and $^{232}$Th-chain decays. Most external backgrounds are rejected by a 3.5\,m fiducial radius cut, which preserves 20\% of signal events. Inside the 3.5\,m fiducial volume (FV) and 2.47\,MeV to 2.70\,MeV energy ROI, the main background sources are:
\begin{description}
\item[$\mathbf{^{8}}$B solar neutrinos:] flat continuum background from the elastically scattered (ES) electrons normalized using the total $^{8}$B flux and published solar mixing parameters \cite{SNO_3ph}.
\item[2$\boldsymbol{\nu\beta\beta}$:] irreducible background due to the $2\nu\beta\beta$ decays of $^{130}$Te: these events appear in the ROI due to the energy resolution of SNO+.
\item[External Backgrounds:] $^{208}$Tl and $^{214}$Bi nuclides contained in the AV, hold-down rope system, water shielding, and PMT glass are the major contributors in the defined ROI. %\cite{whitepaper}. 
The FV cut of 20\% reduces these background events by several orders of magnitude. The PMT hit-time distribution cut reduces the external background events falling in the FV by an additional factor of two.
\item[Internal $^{238}$U- and $^{232}$Th-chain Backgrounds:] the dominant backgrounds in the signal ROI are due to $^{214}$Bi-Po and $^{212}$Bi-Po decays. Currently, we have achieved approximately 100\% rejection of separately triggered $^{214}$Bi-Po and $^{212}$Bi-Po decays falling inside the ROI and FV using the $\beta$-$\alpha$ delayed coincidence. For $^{212}$Bi-Po and $^{214}$Bi-Po pile-up decays, cuts based on PMT hit timing achieve a rejection factor of $\sim$50 for events that fall in the ROI and FV. Other minor contributions in the ROI are due to $^{234m}$Pa ($^{238}$U chain), $^{210}$Tl ($^{238}$U chain) and $^{208}$Tl ($^{232}$Th chain). %\cite{whitepaper}.
\item[Cosmogenic Backgrounds:] The most relevant isotopes are $^{60}$Co, $^{110m}$Ag, $^{88}$Y, and $^{22}$Na (see Section \ref{sec::cosmo}). The developed purification techniques together with a long period of underground storage will reduce the cosmogenically induced background to less than one event per year in the FV and ROI.
\item[($\boldsymbol{\alpha}$,n) Backgrounds:] both the prompt signal and the delayed 2.22\,MeV-$\gamma$ produced by ($\alpha$,n) reactions can leak into the $0\nu\beta\beta$ ROI. Coincidence-based cuts have been developed that remove more than 99.6\% of the prompt and $\sim$90\% of delayed events that fall in the FV and ROI.
\item[Pile-up Backgrounds:] the most important pile-up backgrounds for the $0\nu\beta\beta$ search are due to high-rate $^{210}$Po+$2\nu\beta\beta$ and $^{210}$Bi+$2\nu\beta\beta$, with bismuth and polonium coming from both the TeLS and the vessel surface. Timing-based cuts have been developed that reduce the pile-up backgrounds to a negligible level.
\end{description}

We have estimated the fraction of each background that falls in the ROI and FV based on our Monte Carlo simulations. %\cite{whitepaper}. 
A summary of the various background sources in the ROI and FV is shown in Table \ref{tab::back_dbd}. The main contributions are due to $^{8}$B $\nu$ ES and to $2\nu\beta\beta$. A total of about 22 events/yr in the FV and ROI are expected. The scale of the external background events within the ROI can be checked by fitting events outside the fiducial volume. Internal U- and Th-chain residuals can be checked via the $^{214}$Bi-Po and the $^{212}$Bi-Po delayed coincidences, whose tagging efficiency can be tested during the pure LAB-PPO scintillator phase. In addition, some of the cosmogenic-induced backgrounds, like $^{124}$Sb and $^{88}$Y, can be constrained using their relatively short half-life, while $^{8}$B-$\nu$ and $2\nu\beta\beta$ decays can be constrained by their known value. Furthermore, the detector response will be tested through a detailed calibration (see Section \ref{sec::calib}).

\begin{figure}[t]
\centering
\includegraphics[scale=0.41]{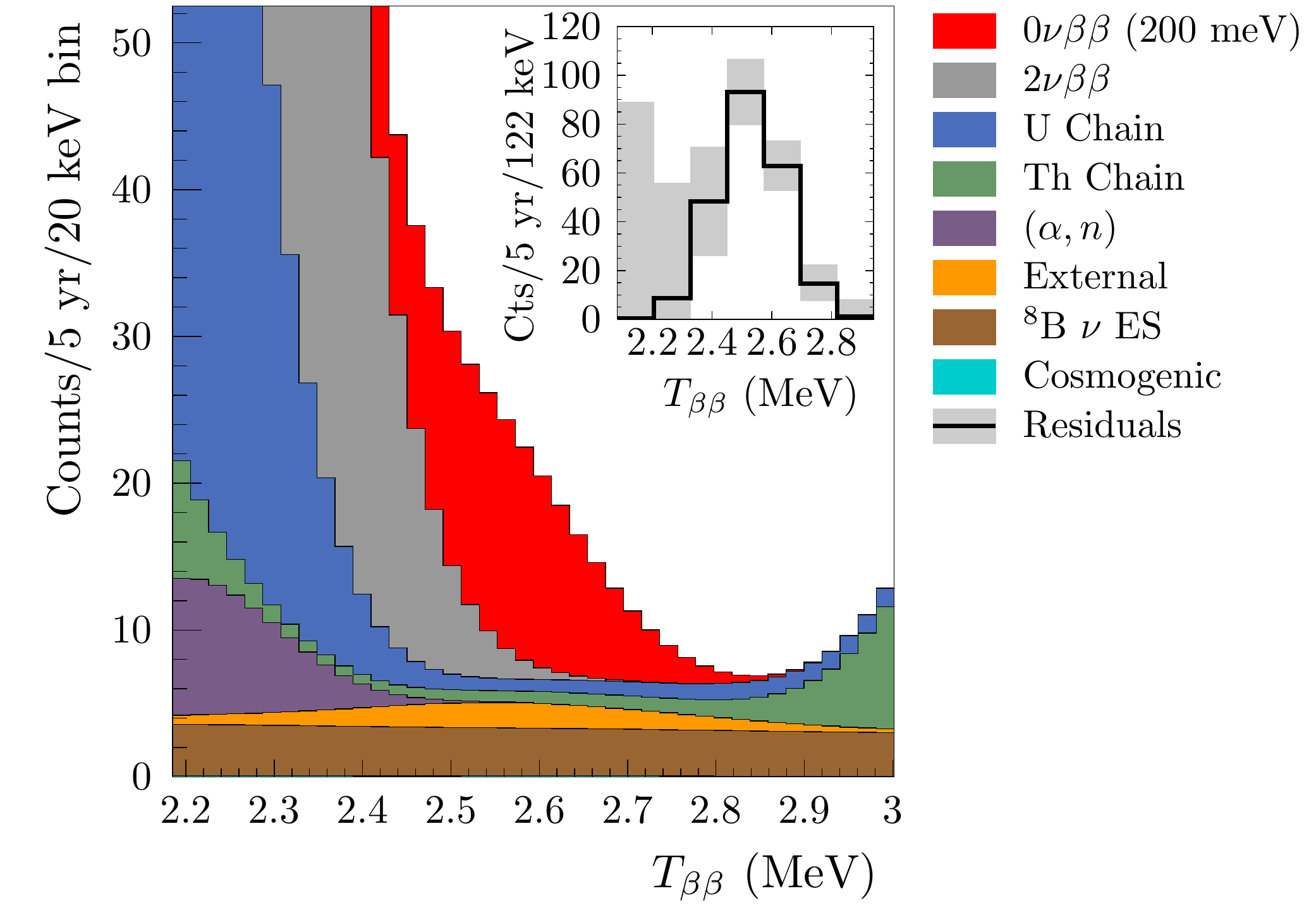} 
\caption{Summary stacked plot of all backgrounds and a hypothetical 0$\nu\beta\beta$ signal corresponding to a mass $m_{\beta\beta}~=~200$\,meV for 5 years data taking. Events are shown in the FV of 3.5\,m, for 0.3\% natural tellurium loading and 200\,Nhits/MeV light yield. T$_{\beta\beta}$ is the effective kinetic energy. \label{fig::Te_en}}
\end{figure}

The expected signal and background spectrum for a five-year live-time is shown in Figure\,\ref{fig::Te_en} for the 0.3\% loading. A fiducial volume cut is applied at 3.5\,m, $>$99.99\% rejection for $^{214}$Bi-Po and $>$98\% for $^{212}$Bi-Po are assumed, and the light yield is 200\,Nhits/MeV. The $0\nu\beta\beta$ signal shown is for a $m_{\beta\beta}$\,=\,200\,meV, which corresponds to $T_{1/2}^{0\nu\beta\beta}\sim1\times10^{25}$ yr using the IBM-2 nuclear matrix element \cite{Barea}.

\begin{table}[t]
\centering
\caption{Expected background counts in the signal ROI and 3.5\,m FV in SNO+ for the first year (Year 1) and in 5 years of the 0.3\% Te loading phase. A light yield of 200\,Nhits/MeV has been assumed. Cuts have been applied as decribed in the text. \label{tab::back_dbd}}
\begin{tabular}{l c c} \hline 
Isotope                 &       1 Year  &       5 Years \\
\hline
$2\nu\beta\beta$        &       6.3   &       31.6  \\
$^{8}$B $\nu$ ES        &       7.3   &       36.3  \\ 
Uranium Chain     &        2.1   &      10.4       \\
Thorium Chain     &   1.7          &  8.7      \\
External                &       3.6   &      18.1  \\
$(\alpha,n)$  & 0.1     &      0.8  \\
Cosmogenics              &       0.7   &       0.8  \\ \hline
Total           &       21.8 &       106.8\\ \hline 
\end{tabular}
\end{table}

%\begin{figure}[t]
%\centering
%\includegraphics[scale=0.45]{figure/pie.pdf} 
%\caption{Summary of all the background sources contributing in the ROI and FV with 0.3\% (by weight) Te loading and 200\,Nhits/MeV. A total of about 22 events/yr is expected.\label{fig::DBDpie}}
%\end{figure}

\subsection{Sensitivity}

The expected number of $0\nu\beta\beta$ events occurring in the SNO+ detector is given by:
\begin{equation}
\label{eq:signal-lifetime}
S = \epsilon \cdot N_{130} \cdot \ln{2} \cdot \frac{t}{T^{0\nu\beta\beta}_{1/2}}
\end{equation}

\noindent where $\epsilon$ is the signal detection efficiency, $N_{130}$ is the number of $^{130}$Te atoms in the detector, $t$ is the live-time, and
$T^{0\nu\beta\beta}_{1/2}$ is the half-life of $^{130}$Te $0\nu\beta\beta$. To compute the SNO+ sensitivity, we assume that the number of observed events in the FV and ROI is equal to the expected backgrounds. In this case the numerical value of the derived bound on the number of signal events is similar for either a Bayesian or a frequentist definition of 90\% confidence level. With the natural tellurium concentration of 0.3\% (by weight) in Phase I, corresponding to about 800\,kg of $^{130}$Te, a 20\% FV cut, and five years of data taking, SNO+ can set a lower limit on the half-life of T$^{0\nu\beta\beta}_{1/2}>$\,9\,$\times10^{25}$\,yr at 90\% CL (T$^{0\nu\beta\beta}_{1/2}>$\,4.8\,$\times10^{25}$\,yr at 3$\sigma$ level). This corresponds to a limit on the effective Majorana neutrino mass, $m_{\beta\beta}$, of $55\,\textrm{--}\,133$\,meV, using a phase space factor $G~=~3.69\times 10^{-14}$\,yr$^{-1}$ \cite{Kotila} and $g_{A}~=~1.269$; the range is due to differences in nuclear matrix element calculation methods \cite{Barea, Sim, ism, pnQRPA, edm}.

\subsection{Higher Tellurium Concentration in the Future}

One of the main advantages of the SNO+ technique is the possibility of moving toward higher sensitivities by increasing the loading. R\&D efforts have demonstrated that, with 3\% (by weight) tellurium loading, a light yield of 150\,Nhits/MeV can be achieved using perylene as a secondary wavelength shifter. In SNO+ Phase II, this loss in light yield will be compensated by an upgrade to high quantum efficiency PMTs and improvements to PMT concentrators. These improvements will increase the light yield by a factor of $\sim3$. A preliminary study shows that SNO+ Phase II can set a lower limit on the $0\nu\beta\beta$ half-life of $T_{1/2}^{0\nu\beta\beta}>7\times10^{26}$ years (90\% CL), for a $m_{\beta\beta}$ range of 19--46\,meV. %\cite{whitepaper}.

\section{Solar Neutrino Physics}\label{sec::solar}

SNO+ has the opportunity to measure low energy solar neutrinos with unprecedented sensitivity. This is due to
the reduced production rate of cosmogenic isotopes at the SNOLAB depth and requires that the intrinsic background sources are low enough.

At scintillator purity levels similar to that of Borexino Phase I~\cite{bxo09, bxo091}, the unloaded scintillator phase of SNO+ provides excellent sensitivity to CNO, \textit{pep}, and low energy $^{8}$B neutrinos. With the scintillator sourced from a supply low in $^{14}$C, SNO+ could also measure {\it pp} neutrinos with a sensitivity of a few percent.  Due to the relatively high end-point of the spectrum, $^{8}$B $\nu$s with energy above the $^{130}$Te end-point can also be measured during the \obb phase.

The first measurement of the flux of neutrinos from the subdominant CNO fusion cycle would constrain the metallicity of the solar interior and thus provide critical input to the so-called solar metallicity problem: the current disagreement between helioseismological observations of the speed of sound and model predictions, due to uncertainties in the heavy element (metal) content of the Sun.  Historically, model predictions for the speed of sound were in excellent agreement with observation, one of the primary reasons for confidence in the Standard Solar Model during the period of uncertainty surrounding the solar neutrino problem.  However, recent improvements in solar atmospheric modeling, including transitioning from one-dimensional to fully three-dimensional models, and including effects such as stratification and inhomogeneities~\cite{asplund}, produced a lower value for the heavy element abundance of the photosphere and, thus, changed the prediction for the speed of sound.  The theoretical prediction for the CNO flux depends linearly on the core metallicity and can be further constrained by a precision measurement of the $^8$B flux, due to their similar dependence on environmental factors.  A measurement of CNO neutrinos would thus resolve this uncertainty and also advance our understanding of heavier mass main-sequence stars, in which the CNO cycle dominates over the {\it pp} fusion chain.

Precision measurements of the {\it pep} flux and the low energy $^8$B spectrum offer a unique opportunity to probe the interaction of neutrinos with matter and to search for new physics. The shape of the $\nu_e$ survival probability in the transition region between vacuum oscillation ($\leq 1$\,MeV) and matter-enhanced oscillation ($\geq 5$\,MeV) is particularly sensitive to new physics effects, such as flavor changing neutral currents or mass-varying neutrinos, due to the resonant nature of the MSW interaction.  %Sensitivity to both {\it pep} and $^8$B neutrinos is a  for this purpose.  
The {\it pep} neutrinos are a line source at 1.44\,MeV, thus offering the potential for a direct disappearance measurement partway into this vacuum-matter transition region.  However, due to their production region closer to the core of the Sun, the effect of  new physics on the $^8$B neutrino spectrum is significantly  more pronounced. Thus, the most powerful search combines a precision measurement of the {\it pep} flux with a  $^8$B spectral measurement.  

Borexino has published the first evidence for {\it pep} neutrinos~\cite{gal11}, with a significance of just over 2$\sigma$ from zero.  In order to distinguish different models, a precision of at least 10\% is required.  A number of experiments have extracted the $^8$B spectrum~\cite{SNO_3ph, KL8B, Bor8B, SK8B, SK88-2}, and there is some weak evidence for non-standard behaviour in the combined data set~\cite{Bonv} but the significance is low (roughly 2$\sigma$).  
The theoretical uncertainty on {\it pep} neutrinos is very small, and well constrained by solar luminosity measurements.  The $^8$B flux is well measured by the SNO experiment \cite{SNO_3ph}.  Precise oscillation measurements are therefore possible.

Should the SNO+ scintillator be sourced from a supply naturally low in $^{14}$C, similar to or within an order of magnitude or so of the level observed in Borexino, there also exists the potential for a precision measurement of {\it pp} neutrinos.  Borexino has produced the first direct detection of these neutrinos, with a precision of a little over 10\%~\cite{bxo14}.  A percent level measurement would allow a test of the so-called luminosity constraint, thus testing for additional energy loss or generation mechanisms in the Sun, and allowing us to monitor the Sun's output using neutrinos.

\subsection{Backgrounds}

%\begin{figure*}[t]
%\centering
%\subfigure[SNO+]{\includegraphics[scale=0.4]{figure/SNO+_11C.pdf} } \hspace{0.5cm}
%\subfigure[Borexino]{\includegraphics[scale=0.4]{figure/BXO_11C.pdf} }
%\caption{Simulated comparison between the expected $^{11}$C background in the SNO+ experiment (extrapolated from KamLAND data \cite{kam10}, a) and in the Borexino experiment \cite{bxo11} (b) as a function of depth and fiducial volume for 1 year of measurement time. The resolution used is 5\%/$\sqrt{E}$ (MeV). Other expected backgrounds are not shown.\label{fig::11c}}
%\end{figure*}

The sensitivity of the SNO+ solar phase will depend critically on the leaching rate of $^{210}$Bi. As described in Section \ref{leaching}, radon daughters, implanted on the internal AV surface, are expected to leach off during the various SNO+ phases with a rate that depends both on the temperature and on the liquid in contact with the vessel. We will be able to evaluate the levels of these backgrounds both during the initial water fill and during the scintillator fill itself. We are also investigating mitigation techniques to be applied in case the background levels are initially too high to perform the solar measurement. These techniques include $in$--$situ$ recirculation, further purification, and the use of a balloon to shield from external backgrounds.

Other backgrounds for the measurements of \textit{pep} and CNO neutrinos are the levels of $^{214}$Bi ($^{238}$U chain), $^{212}$Bi ($^{232}$Th chain), and $^{11}$C in the pure scintillator. $^{238}$U and $^{232}$Th levels in the scintillator can be effectively constrained using the $\beta$-$\alpha$ delayed coincidence, as described in Section \ref{sec::Background}. $^{11}$C decays, which were the main background for the measurement of \textit{pep} neutrinos in Borexino \cite{gal05}, can be identified by a three-fold coincidence algorithm (see Section \ref{sec::cosmo}).

Another muon-induced isotope that is a potential background for low energy $^{8}$B neutrino searches is $^{10}$C (T$_{1/2}$\,=\,19.3\,s, Q-value\,=\,3.65\,MeV). However, due to the isotope's short half-life and the low cosmic muon rate at SNOLAB depth, it can be removed by cutting events that occur within a few minutes from each muon event.

%The levels of $^{14}$C and $^{85}$Kr in the pure scintillator are important background for the measurement of the lowest-energy \textit{pp} solar neutrinos, which have been recently observed by Borexino \cite{bxo14}. 

\subsection{Sensitivity}

Sensitivity studies were performed assuming one year of unloaded scintillator data, which could be either prior to or following the Te-loaded phase.  An extended maximum likelihood fit was performed in energy, with a conservative 50\% fiducial volume, in order to reduce external background contributions to negligible levels.  A two-dimensional fit would allow an increase in fiducial volume and thus improve sensitivity.  Thirty-four signals were included in the fit: the four neutrino signals ($^8$B, $^7$Be, CNO, and {\it pep}) as well as thirty background event types.  Backgrounds expected to be in equilibrium were constrained to a single fit parameter; $^{210}$Po, $^{210}$Pb, and $^{210}$Bi were treated independently, that is, not assumed to be in equilibrium with the parent decays.  Background parameters included in the fit were the normalisations of: $^7$Be, $^{39}$Ar, $^{40}$K, $^{85}$Kr, $^{210}$Po, $^{210}$Pb, $^{14}$C, $^{238}$U chain, and $^{232}$Th chain.  $^{210}$Bi was linked to CNO in the fit due to the similarity of the energy spectra; the separation is best achieved by imposing an {\it ex--situ} constraint on the level of $^{210}$Bi decays, or by using observables other than energy.

The nominal background levels assumed were those achieved by Borexino during their initial running.  It was assumed that purification techniques (in particular, distillation) can reduce $^7$Be contamination to negligible levels.  Gaussian constraints were applied to backgrounds where an {\it ex--situ} or independent {\it in--situ} measurement of the rate is anticipated. $\alpha$ tagging is expected to reduce the $^{210}$Po peak by 95\%, with an uncertainty of 20\% on the remaining 5\% of the events. Coincidence decays provide a 50\% constraint on $^{85}$Kr, 25\% on the $^{232}$Th-chain backgrounds, and 7\% on the portion of the $^{238}$U chain that is treated as being in equilibrium.

The fit range was between 0.2\,MeV and 6.5\,MeV, with 10\,keV bins in visible energy.  Extending the fit to higher energies would improve the accuracy on the $^8$B flux measurement.  Bias and pull tests show that the fit is stable and accurate, and robust to changes in bin size or energy range (to within changes in statistics, e.g. $^8$B flux accuracy is reduced if the energy range of the fit is reduced).

\begin{figure}[t]
\centering
\includegraphics[scale=0.42]{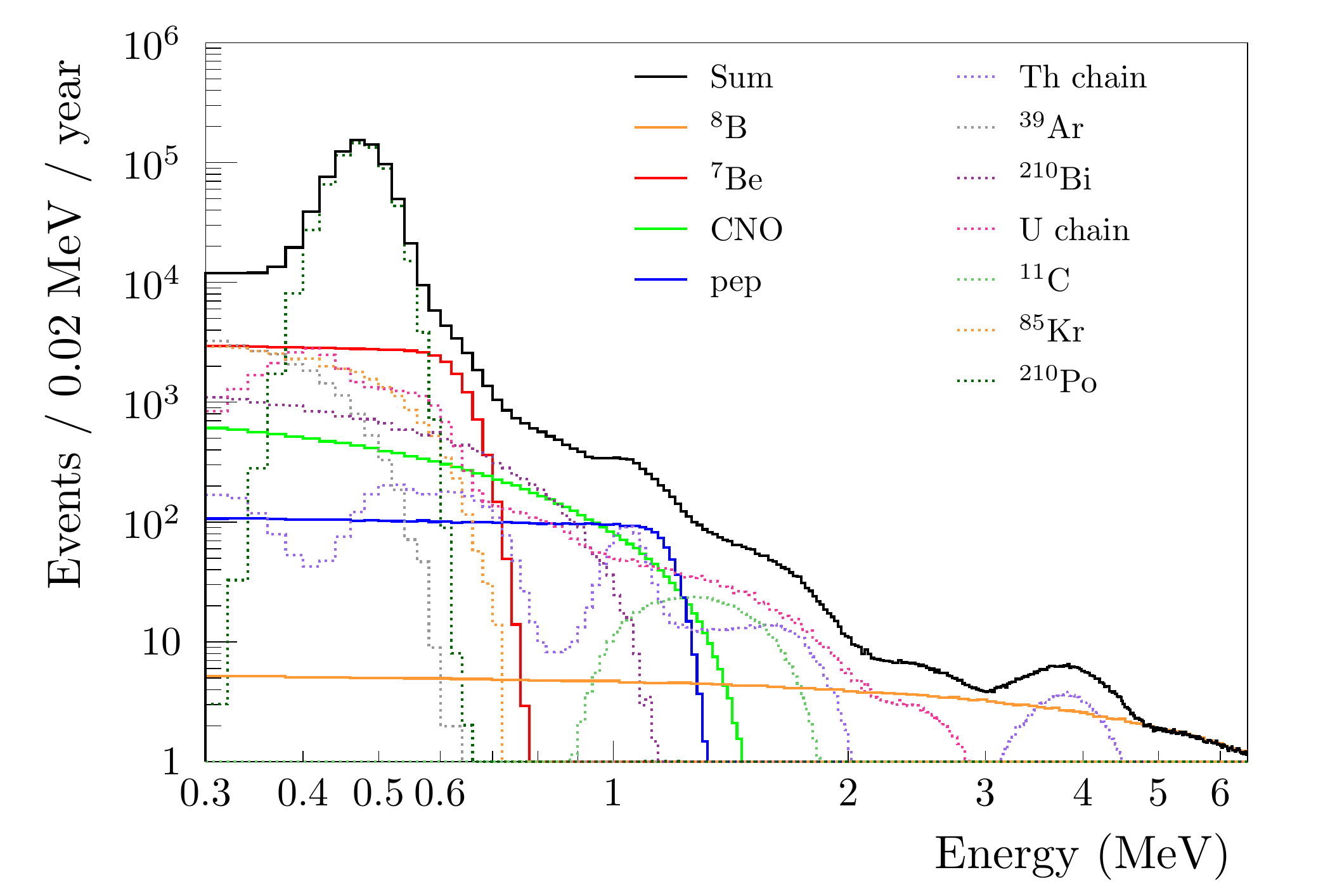} 
\caption{Expected solar neutrino fluxes as detected by SNO+ and the corresponding main backgrounds. Backgrounds levels are assumed to be equal to those initially achieved by Borexino \cite{bxo09, bxo091} (see text). Events are shown for the LAB-PPO scintillator, 400\,Nhits/MeV light yield, and a fiducial volume cut of 5.5\,m. A 95\% reduction is applied to the $^{214}$Bi-Po backgrounds via delayed coincidence tagging, and a 95\% reduction on the $^{210}$Po and the remaining $^{214}$Po events via alpha tagging. \label{fig::solar}}
\end{figure}

The simulations suggest that, with one year of data, the uncertainty on the {\it pep} flux will be less than 10\%.  The uncertainty on the linked CNO+$^{210}$Bi flux is 4.5\%, into which we fold a conservative uncertainty for separating the two signals, resulting in a 15\% predicted uncertainty on the CNO flux.  The $^7$Be flux can be measured to 4\%, and $^8$B to better than 8\%. The uncertainty on the neutrino flux measurements is dominated by statistics, and by correlations between the neutrino signals themselves.  A study of energy scale and resolution systematics shows that these parameters can be floated as nuisance parameters in the fit, and the data will constrain them to better than the required precision, with sub-percent level impact on the neutrino flux uncertainties.  Calibration sources will be deployed in order to measure effects such as any non-Gaussianity of the resolution function, and any potential nonlinearity in the energy scale. In Figure\,\ref{fig::solar} the full solar neutrino signals as detected by SNO+ are shown together with the main background sources for the LAB-PPO scintillator. A fiducial volume cut is applied at 5.5\,m.

$^{214}$Bi-$^{214}$Po events are reduced by 95\% using the $\beta$-$\alpha$ delayed coincidence as described in Section \ref{sec::238U}. A 95\% rejection is applied to the $^{210}$Po events and the remaining $^{214}$Po events via alpha tagging. There is no rejection applied to the $^{212}$Bi and $^{212}$Po events. This is a conservative approach as we expect to reject the majority of these events using a $\beta$-$\alpha$ delayed coincidence as for the $0\nu\beta\beta$ search (see Section \ref{sec::doublebeta}).

Studies show that the precision with which the {\it pp} neutrinos could be observed depends critically on the levels of backgrounds such as $^{14}$C and $^{85}$Kr in the scintillator.  If these backgrounds are low, within 10--50 times that seen in Borexino, SNO+ could achieve a few-percent level measurement of the {\it pp} neutrino flux with just 6 months of solar neutrino data.

\section{Antineutrino Studies}\label{sec::antinu}

Antineutrino events in SNO+ will include geoneutrinos from the Earth's radioactive chains of uranium and thorium, antineutrinos from nuclear reactors, and the antineutrinos emitted by a supernova burst (which are considered in detail in Section \ref{sec::supernova}). The measurement of geoneutrinos will constrain the radiogenic heat flow of the Earth for geophysics studies, while the measurement of reactor antineutrinos, with a known energy spectrum and a precise propagation distance, can better constrain the neutrino oscillation parameters \cite{kamlandreactor}. 

\subsection{Signal Detection}
Antineutrinos are detected in SNO+ via inverse beta decay (IBD): $\overline{\nu}_{e}$s with energy greater than 1.8\,MeV interact with the protons in the liquid scintillator, producing a positron and a neutron. The antineutrino energy is measured by the scintillation light emitted by the positron as it slows down and annihilates:
\begin{equation}
E_{\bar{\nu}_e} \simeq E_{prompt} + (M_n - M_p) - m_e \simeq E_{prompt} + 0.8\,\rm{MeV}
\end{equation}
where M$_n$, M$_p$, and m$_e$ are the neutron, proton, and electron masses. The neutron emitted in the reaction will first thermalize and then be captured by hydrogen, leading to the characteristic 2.22\,MeV delayed gamma from the deuterium formation. The prompt+delayed signal allows the identification of the antineutrino event. The coincidence time interval is defined by the period elapsed from neutron emission to its capture, generally about 200\,$\mu$s, while the spatial separation between the prompt and the delayed event depends on the distance travelled by the delayed gamma before scintillation light is emitted.
The exact values to use for the time and distance coincidence tag, to identify the $\overline{\nu}_{e}$ events, depend crucially on the correct simulation of the neutron propagation in the scintillator mixture being used (unloaded or Te-loaded scintillator). Neutron propagation in each of the scintillator cocktails planned by SNO+ will be checked with a detailed calibration program using an AmBe source. This source, already extensively used by SNO, has a well-known neutron energy spectrum, extending to energies higher than those of the expected antineutrino signals. The calibration results will be cross-checked with a detailed Monte Carlo simulation.

\subsection{Backgrounds}
As the antineutrino signal is identified as a delayed coincidence in SNO+, the main backgrounds are true or random coincidences in the detector with the identified neutron capture. Most of the background neutrons are expected to come from external background sources and are therefore captured and reconstructed in the external regions of the detector. Events that reach the region inside the vessel can be mitigated by a fiducial volume cut, or by a radius-dependent analysis. The major source of neutrons inside the scintillator is the $(\alpha$,n) reactions, which are mainly caused by $^{210}$Po-alpha leached off the vessel surface and are expected to increase with time, as described in Section \ref{sec::an}. The associated prompt signal, mainly due to the proton recoil, will be at energies lower than 3.5\,MeV or, in case the product nucleus is in an excited state, in definite gamma peaks which will allow the study of the ($\alpha$,n) background's time evolution.

\subsection{Reactor Antineutrinos and Oscillations}\label{reactor}

\begin{figure}[t]
\centering
\includegraphics[width=3.5in]{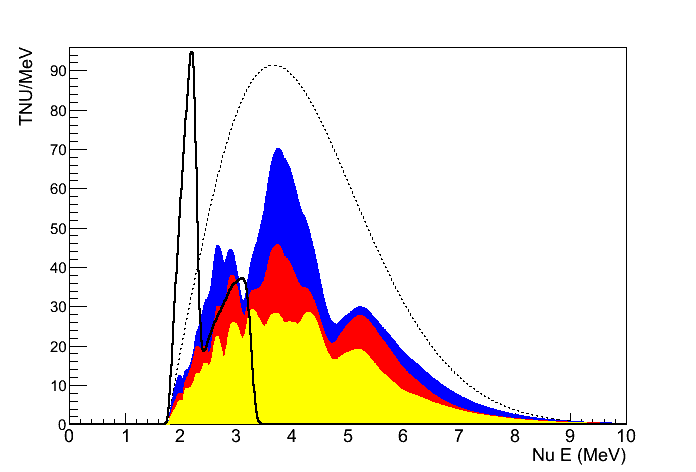}
\caption{Expected visible antineutrino energy spectrum in SNO+, for 10$^{32}$ proton-years per MeV. The nonoscillated reactor spectrum (dashed line) is shown together with the geoneutrino spectrum (solid line, arbitrary normalization). The stacked oscillated reactor spectrum is shown with different colors, each corresponding to a reactor complex: reactor at 240\,km in blue (top), reactors at 350\,km in red (middle), and other reactors in yellow (bottom). See text for details. \label{fig::antinu}} 
\end{figure}

In SNO+ we expect around 90 reactor antineutrino events per year. The total flux is obtained summing 3 components: (1) 40\% of it comes from one reactor complex in Canada at a baseline of 240\,km, (2) 20\% is from two other complexes at baselines of around 350\,km, and (3) 40\% is divided between reactors in the USA and elsewhere at longer baselines. The signals from the first two sources (1 and 2) induce a very clear oscillation pattern (see Figure\,\ref{fig::antinu}), which lead to a high sensitivity to the $\Delta m^2_{12}$ neutrino oscillation parameter. For E\,$<$\,3.5\,MeV the geoneutrino signals and reactor signals overlap. Most of the backgrounds are concentrated in the energy region of the geoneutrinos. For a preliminary study of the reactor neutrino oscillation sensitivity, we conservatively exclude the region below 3.5\,MeV. Assuming a light yield of 300 Nhits/MeV, expected for the Te-loaded phase with perylene as secondary wavelength shifter, and a 5.5\,m FV cut, we expect to reach a sensitivity in $\Delta m^2_{12}$ of 0.2$\times10^{-5}$\,eV$^{2}$, similar to the KamLAND result \cite{kamlandreactor} in about 7 years of data taking. The full analysis will take the complete antineutrino spectrum into account, using constraints for the backgrounds, and measuring simultaneously the geoneutrino flux. 
%Preliminary studies show that in five years of data-taking with unloaded scintillator, using the full antineutrino energy spectrum and assuming full reactor power, SNO+ can measure $\Delta m^2_{12}$ at the same level as KamLAND \cite{kamlandreactor}. We expect to keep this sensitivity also during the Te-loading phase, despite the degradation in energy resolution due to the lower light yield.
%The total normalization, constant in time, is extracted together with the oscillation parameters from the reactor spectrum, as the shape of the irreducible geo-neutrino background for energies up to 3.5\,MeV is known.

Generally, the $\overline{\nu}_{e}$ flux from the Canadian reactors (CANDU-type) is expected to be stable in time due to the continuous refuelling process. However, in the next few years there are expected upgrades in which different reactor cores will be turned off, with only one reactor core switched off at a time in each of the complexes. This will cause changes in the reactor spectrum, with an expected total flux reduction below 10\% at each moment.
This time evolution can be used to identify the very clear oscillation pattern in the reactor spectrum for each of the two identified baselines (240\,km and 350\,km) and to distinguish them from other antineutrino sources.

The oscillation patterns from the more distant reactors are less evident after they are combined. There is still a visible feature at antineutrino energies of 4.5\,MeV from an accumulation of reactors at distances of the order of 550\,km. A detailed description of the spectrum at this energy is still under discussion \cite{DayaBay}. A preliminary study shows that the combined systematic uncertainties associated with the unoscillated spectrum description are below 5\%. These uncertainties can be reduced using, for the distant reactors (source 3), direct measurements at close-by detectors, like those of Daya Bay \cite{DayaBay}.
%The study of how this uncertainty can be best extrapolated to the CANDU reactors is still on-going.

\subsection{Geoneutrinos and Earth Studies}

Interest in geoneutrinos has increased in the last few years with significant collaborations between neutrino physicists and geo-physicists. Joint results may finally explain the radiogenic heat flow of the Earth.

In SNO+ the geoneutrinos from the uranium and thorium chains can be detected. These antineutrinos come mainly from thick continental crust, with increases due to variations in local crust components \cite{sno+geo}. 

The energy spectra of geoneutrinos are well-known for each of the standard decay chains \cite{geo1}. The effect of neutrino oscillations is largely averaged out due to the long range in production distances, leading to a total survival probability of:
\begin{equation}
\langle P_{ee} \rangle=\cos^4\theta_{13}\cdot \left(1-\frac{\sin^2(2\theta_{12})}{2}\right) + \sin^4\theta_{13} \simeq 0.547
\end{equation}
where $\theta_{13}$ = 9.1\degree and $\theta_{12}$ = 33.6\degree \cite{SNO_3ph}.
Detailed studies of the impact of the MSW effects on the energy spectrum are in progress.

As a first analysis step, we will fix the total U/Th ratio according to standard geological models \cite{geo3}, and fit for the total flux assuming a precise shape for the energy spectrum of geoneutrinos. The possible effect of local variations of this ratio is being quantified together with that from the low energy reactor spectrum. Systematic uncertainties in the energy scale and energy resolution and from the constraints on the alpha-n backgrounds will vary for each of the data taking phases. Overall, the SNO+ sensitivity to the total flux is expected to be dominated by statistical uncertainties. The accuracy will be close to that of Borexino for similar data-taking periods: the larger volume of the SNO+ detector compensates for the higher rate reactor background. We expect a similar rate of geoneutrinos and reactor antineutrinos in the 1.8\,MeV--3.5\,MeV energy region. However, the reactor spectrum extends up to much higher energies and contains features that can help in establishing the oscillation parameters. The time evolution analysis will also help to separate the reactor background (Section \ref{reactor}). In the Te-loaded phase the low energy backgrounds are expected to be about 50--150 times higher than in the pure scintillator phase, which can make the extraction of the geoneutrino signal more difficult.

We aim to additionally separate both the uranium and thorium contributions and the mantle and crust contributions in a global analysis of the geoneutrino spectrum including data from KamLAND \cite{kamlandgeo} and Borexino \cite{borexinogeo}. 

\section{Supernova Neutrino Observation}\label{sec::supernova}

The era of neutrino astronomy commenced with the observation of 24 events, all associated with the inverse beta decay of $\bar{\nu}_e$, from the collapse of supernova SN\,1987A at $\sim50$\,kpc \cite{kri04}. SNO+, with its large high purity liquid scintillator volume and the deep location underground, is one of the most promising experiments for the detection of neutrinos from core collapse supernovae (CCSNe), offering a rich sample of detection channels, low backgrounds, and a large number of target particles and nuclei. CCSNe are an exceptional source of neutrinos of all flavors and types, and a measurement is expected to shed light on the explosion mechanism.
%\cite{hir88, hai88, ale88}
The shape of the individual supernova (SN) $\nu_\alpha$ ($\nu_\alpha = \nu_e,\, \bar{\nu}_e,\, \nu_x$, where in this context $\nu_x$ is the sum of $\nu_\mu$, $\overline{\nu}_{\mu}$, $\nu_\tau$ and $\overline{\nu}_{\tau}$) energy spectra is expected to approximate a thermal spectrum \cite{kei03} in the absence of neutrino flavor changing mechanisms. At postbounce times t~$<$~1\,s, before shock revival, the flavor changes are expected to be reduced to those induced by the well--known MSW effect in a quasi--static environment \cite{xu13, sar12}. At later times, many further effects interfere, significantly modifying the spectral shape. These effects are nontrivial and still lack a full understanding and a consistent analytical treatment. At present, sensitivity studies to thermal spectral parameters are only meaningful for at most the first second of the burst. It is estimated that half of all neutrinos are emitted in this time span \cite{pag09}.

\subsection{Signal Detection in SNO+}

\begin{table}[t]
\begin{minipage}{8.5cm}
\begin{center}
\caption[Supernova neutrino interaction channels in liquid scintillator]{\label{tab:SNchannels} Supernova neutrino interaction channels in LAB-PPO. The event rates, per 780 tonnes of material, assume the incoming neutrino time-integrated flux described in the text. No flavor changing mechanisms are considered. The uncertainties on the event rates only include the cross section uncertainties \cite{vK3}.}
\begin{tabular}{p{5.7cm} c}
\hline\noalign{\smallskip}
Reaction & Number of Events \\ \hline
NC: $\nu + p \rightarrow \nu + p$ & $429.1\pm12.0$ \footnote{$118.9\pm3.4$ above a trigger threshold of 0.2\,MeV visible energy.}  \\ \hline
CC: $\bar{\nu}_e + p \rightarrow n +e^+$ & $194.7\pm1.0$ \\\hline
CC: $\bar{\nu}_e$ + $^{12}$C $\rightarrow$ $^{12}$B$_{g.s.}$ + e$^+$ & $7.0\pm0.7$ \\
CC: $\nu_e$ + $^{12}$C $\rightarrow$ $^{12}$N$_{g.s.}$ + e$^-$ & $2.7\pm0.3$ \\ % 3.803
NC: $\nu$ + $^{12}$C $\rightarrow$ $^{12}$C$^*$(15.1\,MeV) + $\nu\,'$ & $43.8\pm 8.7$ \\\hline
CC/NC: $\nu$ + $^{12}$C $\rightarrow$ $^{11}$C or $^{11}$B + X & $2.4\pm0.5$ \\\hline
$\nu$--electron elastic scattering & 13.1\footnote{\label{noteSN}The Standard Model cross section uncertainty is $<1\%$.}\\ \hline
\end{tabular}
\end{center}
\end{minipage}
\end{table}

%\begin{figure*}[ht]
%	\centering
%	\subfigure[True proton recoil spectrum.]{\label{fig:nupES:1}\includegraphics[width=0.45\textwidth]{supernova/plots/SN_vpES_true_thesis.eps}}\hspace{0.5cm}
%	\subfigure[Observed proton recoil spectrum.]{\label{fig:nupES:2}\includegraphics[width=0.45\textwidth]{supernova/plots/SN_vpES_qud_thesis.eps}}
%	\caption{\label{fig:nupES} True (a) and observed (b) proton recoil spectrum in LAB--based scintillator resulting from the reference SN. The observed proton spectrum takes ionization quenching with $kB=0.0098$\,cm/MeV and an energy resolution of $5\%/\sqrt{E_{\mathrm{vis}} \mathrm{(MeV)}}$ into account, where the latter detector effect has no visible impact in this case.}
%\end{figure*}

%For a thermal energy distribution of the neutrinos emitted in a SN explosion \cite{kei03}, the time--integrated flux, i.e. the fluence, on Earth is typically described by \cite{bea11}:
%
%\begin{equation}
%\label{equ:dNdE}
%\frac{d\Phi_\alpha}{dE} = \frac{2.35 \times 10^{13}}{\mathrm{cm^2\,MeV}} \sum \limits_{\alpha} \frac{\varepsilon_{\alpha}}{d^2} \frac{E^3}{\langle E_\alpha\rangle^5} \mathrm{exp}\left(- \frac{4E}{\langle E_\alpha \rangle}\right),
%\end{equation}
%
%where $\langle E_\alpha \rangle$ is the mean neutrino energy and $\varepsilon_\alpha$ is the total SN neutrino energy in units of foe (1\,foe\,$=10^{51}$\,erg). 

For the detection potential of SNO+ presented in this paper, we assume that the distance from the SN to Earth is $d~=~10$\,kpc -- known from, for example, the detection of the electromagnetic radiation released in the SN event, and that 3$\times10^{53}$ erg of binding energy ($\epsilon_\nu$) are released in the form of neutrinos, equally partitioned amongst all six flavors and types. The mean energies used are 12\,MeV for $\nu_e$, 15\,MeV for $\bar{\nu}_e$ and 18\,MeV for $\nu_x$ \cite{bea11}, which are generic mean SN neutrino energies \cite{lun13} consistent with the findings from SN\,1987A.

The possible SN neutrino interaction channels during the SNO+ pure scintillator phase are listed in Table\,\ref{tab:SNchannels} together with the expected event rates. Several events due to \bnel\ are expected, because of the comparatively large cross section for the IBD reaction \cite{sch12a}. This process, seen during SN\,1987A, is the only interaction of SN neutrinos observed to date. Additionally, SNO+ can measure the flux of \mbox{$\nu_x$} and \nel. As the mean neutrino energy is below about 30\,MeV, $\nu_e$s and $\overline{\nu}_e$s will be detected mainly by the charged current (CC) interactions, while supernova $\nu_x$s can only be detected by the more challenging neutral current (NC) reactions. One NC reaction is neutrino--proton elastic scattering (ES), $\nu + p \rightarrow \nu + p$ \cite{bea02}, which is the only channel that provides spectral information about the \nx s. The total cross section of this process \cite{ahr87} is about a factor of three smaller than the cross section of IBD; however, the reaction is possible for all six neutrino types yielding a similar number of events for a detector threshold above $\sim$0.2\,MeV.

%The true, $E_p$, and quenched, observed $E_{\mathrm{vis}}$, proton energy distributions resulting from the reference SN fluence are shown in Figure \ref{fig:nupES}, where it is clear that most of the scattering events are shifted below $\sim500$\,keV electron--equivalent energy. 

\subsection{SNO+ Sensitivity to the $\nu_x$ Spectral Shape}

%\begin{figure}[t]
%\centering
%\includegraphics[scale=0.45]{figure/SNOp_Emean_Norm_95CL_thesis.eps} 
%\caption{\label{fig::Contour} 95\% C.L. contour plot for \vp~ signal in SNO+ with and without systematic uncertainties \cite{vK3}. It is clear that statistical uncertainties dominate (see text).}
%\end{figure}
In the preliminary estimations of the SNO+ sensitivity to \nx\, spectral shape through \vp~we conservatively assume a spatial radius cut of 5\,m and a 0.2\,MeV threshold, corresponding to a minimal neutrino energy of $E_\nu^{\mathrm{min}}\approx21.9$\,MeV. This is close to the threshold we expect to use for events that will be permanently stored. We are currently discussing other settings for the trigger thresholds to avoid any loss of potential low energy supernova events.

\begin{figure}[t]
\centering
\includegraphics[scale=0.4]{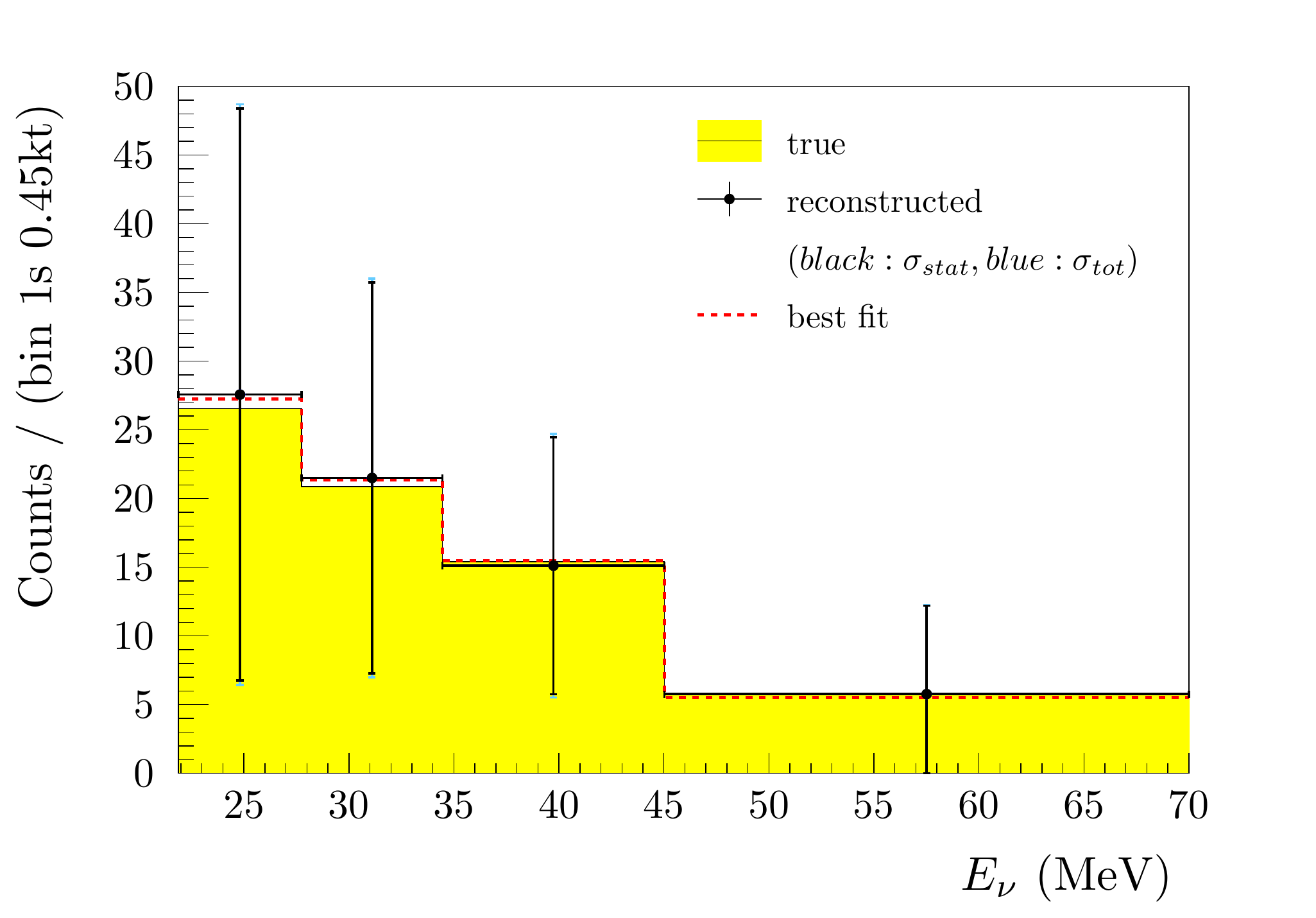} 
\caption{\label{fig::EnuSpec} True, reconstructed, and best fit SN neutrino energy distribution of the \vp\, detection channel within the FV and above the detector threshold \cite{vK3}. Shown are the sum of the $\nu_e$, \bnel\, and \nx\,  spectra, considering their time-integrated flux in the first second of the reference SN. The statistical uncertainties are shown in black, while the total uncertainties are shown in blue. The contribution from systematic uncertainty is too small to be resolved.}
\end{figure}

In Figure\,\ref{fig::EnuSpec} the reconstructed energy spectrum of all neutrinos emitted in the first second of the SN ($\nu_e$, \bnel\, and $\nu_x$) and detected in SNO+ via the \vp~reaction is shown together with the true neutrino spectrum. The reconstructed energy spectrum is obtained from the detected proton energy unfolded using the TUnfold algorithm \cite{tunfold}, on the basis of binned data. The strongly nonlinear quenching of the proton energy, which shifts most of the scattering events below $\sim0.5$\,MeV electron--equivalent energy, and the finite detector resolution are taken into account. The number of events in the lowest bin is slightly overestimated, due to bin--to--bin migrations caused by the finite energy resolution. The statistical and total systematic uncertainties are also shown. A fit to the $\nu_x$ spectrum is only possible if the \nel\, and \bnel\, spectra are measured independently. SNO+ is sensitive to the spectral shape of  $\overline{\nu}_{e}$s via the IBD reaction, while in the case of $\nu_{e}$s it has to be assumed that an independent detector, with, for example, a Pb target like HALO \cite{vaa11} or a LAr target \cite{sch12a}, provides the necessary spectral information. 

The resulting best fit $E_\nu$ spectrum is also shown in Figure\,\ref{fig::EnuSpec} and is in excellent agreement. The systematic uncertainties propagated within the fit are the $\nu$-p ES cross section, the number of target protons, $N_p$, the ionization quenching parameter, the spectral \nel\, and \bnel\, parameters, and the energy resolution of the detector. The corresponding best fit values are $\langle{E}_{\nu_x}\rangle = 17.8^{+3.5}_{-3.0}(\mathrm{stat.})^{+0.2}_{-0.8}(\mathrm{syst.})$\,MeV and $\varepsilon_{\nu_x} = (102.5^{+82.3}_{-42.2}(\mathrm{stat.})^{+16.2}_{-13.0}(\mathrm{syst.}))\times10^{51}$\,erg \cite{vK3}, while the respective expectation values are 18\,MeV and 100$\times 10^{51}$\,erg.

\subsection{SNEWS}

SNO+ is preparing to participate in the inter--experiment Supernova Early Warning System (SNEWS) \cite{ant04}, which has the goal to provide a fast and reliable alert using the coincident observation of burst signals in several operating detectors. As neutrinos escape from the SN tens of minutes up to several hours before the first photons, their detection offers the possibility of alerting the astronomical community to the appearance of the next SN light signal.

\section{Exotic Physics Searches}\label{sec::exotic}

Due to its location deep underground, which significantly reduces the cosmogenic background, and the high radio-purity of the materials used, SNO+ has a unique sensitivity to search for exotic physics, including certain modes of nucleon decay and axion or axion-like particle searches.

\subsection{Invisible Nucleon Decay}\label{sec::nucleon_decay}

Nucleon decay modes to a final state undetected by the experiment, for example, n$\rightarrow 3\nu$, can be searched for by detecting the decay products of the remaining unstable nucleus as it deexcites. This process has been previously investigated by some experiments such as SNO \cite{sno04} by searching for the decay of $^{16}$O nuclei, and Borexino \cite{bxo03} and KamLAND \cite{kam06} by looking for the decay of $^{12}$C nuclei.
We plan to search for the invisible nucleon decay of $^{16}$O during the initial water phase of the experiment. In the case of a decaying neutron, the resulting $^{15}$O will deexcite emitting a 6.18\,MeV gamma 44\% of the time. For a decaying proton, the nucleus is left as $^{15}$N which in 41\% of the decays de-excites emitting a 6.32\,MeV gamma \cite{Eij93}. Both these signals will be in a favorable region of the SNO+ energy spectrum (5.4\,MeV--9\,MeV) in which few backgrounds are expected. These are: (1) internal and external $^{208}$Tl and $^{214}$Bi decays, (2) solar neutrinos, and (3) reactor and atmospheric antineutrinos. The expected contribution of each background in the 5.4--9\,MeV energy region, in six months of running, is shown in Table\,\ref{tab::back_nd}. The targeted purity for the SNO+ internal water is the average of the SNO collaboration's H$_{2}$O and D$_{2}$O levels (see Table\,\ref{tab::back_sources_int}). The purity can be measured \textit{in situ} using events below 5\,MeV and cross checked using water assays. Solar neutrino events can be reduced by placing a cut on the direction of the event, which is reconstructed using the topology of the detected Cherenkov light. Reactor antineutrino events can be tagged using a delayed coincidence. The background due to atmospheric neutrinos is expected to be small based on SNO data \cite{sno04}.

\begin{table}[t]
\caption{Expected backgrounds in the 5.4--9\,MeV energy region during six months of water fill. A fiducial volume cut of 5.5\,m is applied to all events. The events after the $\cos\theta_{sun}> - 0.8$ cut are also shown. $\epsilon$(n) and $\epsilon$(p) are the neutron and proton decay-mode detection efficiencies in the 5.5\,m FV and energy window.\label{tab::back_nd}}
\begin{center}
\scalebox{1.0}{
\begin{tabular}{c c c}
\hline 
Decay source & \multicolumn{2}{c}{Events in six months} \\ 
&  & $\cos\theta_{sun}> - 0.8$ Cut\\ \hline
$^{214}$Bi & 0 & 0 \\ 
$^{208}$Tl & 0.6 & 0.6\\ 
Solar-neutrinos & 86.4 & 17.7 \\ 
Reactor antineutrinos & 1.5 & 1.3\\ 
External $^{214}$Bi-$^{208}$Tl & 9.2 & 8.9 \\ \hline
Total & 97.7& 28.5 \\ \hline
$\epsilon$(n) & 0.1089 & 0.1017\\
$\epsilon$(p) & 0.1264 & 0.1129\\\hline
\end{tabular}}
\end{center}
\end{table}

\begin{figure}[t]
\centering
\includegraphics[scale=0.42]{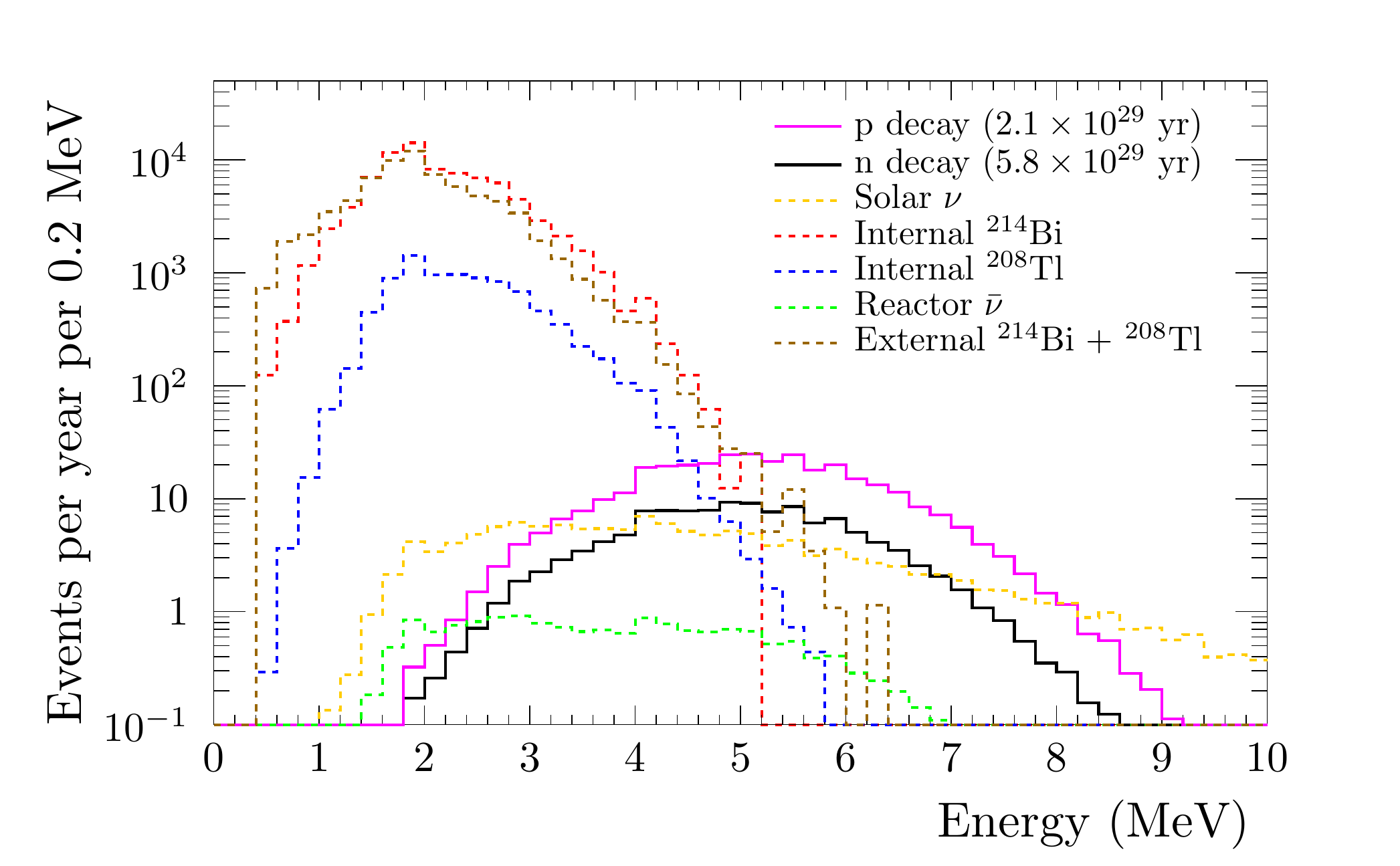}
%\resizebox{0.5\textwidth}{!}{\input{figure/Spectrum_RAT5}}
\caption{Expected energy spectrum for the water phase backgrounds. The signal from invisible proton \cite{sno04} and neutron \cite{kam06} decay is also shown. A fiducial radius cut of 5.5\,m and a cut on $\cos\theta_{sun}> - 0.8$ are applied. \label{fig::nucleon}}
\end{figure}

The events in Table\,\ref{tab::back_nd} are given for a fiducial volume cut of 5.5\,m, which helps in reducing the external backgrounds. An additional cut at cos$\theta_{sun}>-0.8$ relative to the solar direction further reduces the dominant solar background, removing $\sim$80\% of the events with a sacrifice of $\sim$10\% on the signals and the isotropic backgrounds. Figure\,\ref{fig::nucleon} shows the energy spectrum of the water phase backgrounds: solar neutrinos, reactor antineutrinos, and radioactive decays from the uranium and thorium chains, after the two cuts are applied. It also shows the shapes based on the current best limits of the signal gammas from invisible proton \cite{sno04} and neutron \cite{kam06} decay.

Using a Poisson method \cite{Ohel83} we can set the lower limit, at 90\% C.L., on the invisible nucleon decay lifetime $\tau$ by: 
\begin{equation}
\tau > \frac{N_{nucleons} \times \epsilon \times f_{T}}{S_{90\%}}
\end{equation}
\noindent where N$_{nucleons} = 2.4 \times 10^{32}$, $\epsilon$ is the efficiency of detecting the decay in the signal window from Table\,\ref{tab::back_nd}, \textit{S$_{90\%}$} is the expected signal events at 90\% C.L., and $f_{T}$ is the live-time of 0.5 years.
Assuming we reach the expected background, a limit of $\tau_n >$1.25$\times10^{30}$ and $\tau_p >$1.38$\times10^{30}$ years for the decay of neutrons and protons, respectively, can be set. This is an improvement over the existing limit set by KamLAND, $\tau>$ 5.8$\times10^{29}$ years, by a factor of $\sim$2 with just six months of running time. A likelihood approach is in development which is expected to provide a further
improvement on the limit.

\subsection{Axion-Like Particle Search}

An axion-like particle (ALP) is defined as a neutral pseudoscalar particle that exists as an extension to the QCD Lagrangian \cite{axion}. %The deep location underground, the low background, the large volume and the small reactor antineutrino flux, give SNO+ the opportunity to set an improved limit on the ALP coupling constants. 

A possible reaction channel for ALP production in the Sun is p+d$\rightarrow ^{3}$He+A, where A is the ALP with an energy of 5.5\,MeV \cite{bxo12}. In SNO+ the couplings of ALPs to electrons, $g_{Ae}$, photons, $g_{A\gamma}$, and nucleons, $g_{AN}$, can be observed mainly through Compton conversion (A+e$^{-}\rightarrow e^{-}+\gamma$) and the axioelectric effect (A + e$^{-}$+ Z$_{X}$ $\rightarrow$ e$^{-}$  + Z$_{X}$, with \textit{Z$_{X}$} the charge of the involved nucleus \textit{X}). In both cases, for low ALP masses the signature is monoenergetic at $\sim$5\,MeV electromagnetic energy deposition.

Different strategies for different phases of SNO+ are used for the detection of ALPs. In the water phase, the most likely interaction is the Compton conversion, which produces a Cherenkov ring with topology similar to that of $^{8}$B-neutrinos. The main background events are very similar to those described for the invisible nucleon decay search (see Section \ref{sec::nucleon_decay}) as the two signals have similar energies. However, since the Compton conversion has a strong directional bias, we expect to remove a significant amount of isotropic backgrounds, leaving $^{8}$B-neutrinos as the dominant one. With 6 months of water data, due to the deeper location and larger fiducial volume, we expect to approach the current limit set by Borexino \cite{bxo12}.

%During the scintillator phase, due to the much higher light yield than in water and thus the significantly better energy resolution, we expect to improve upon the Borexino limit \cite{bxo12} up to a factor of $\sim$ 5 with three years of data taking.

The BGO collaboration proposed a separate limit on the ALP couplings without having to assume axions interact via Compton conversion \cite{bgo14}. In this case, the detection of solar ALPs via the axioelectric effect, which depends on the nucleus charge as Z$_{X}^{5}$, could be particularly interesting during the Te-loaded phase. Due to the significantly large tellurium mass, SNO+ has the possibility of improving the limit on the axion-electron coupling constant set by the BGO collaboration by several orders of magnitude.

\section{Conclusions}

In this paper the broad physics program of the SNO+ experiment is presented. Three main data taking phases are planned: one with the detector filled with ultra-pure water, one with unloaded liquid scintillator, and one with 2.34 tonnes of tellurium loaded into the detector.

The primary physics goal of SNO+ is a sensitive search for 0$\nu\beta\beta$-decay of $^{130}$Te. We expect to set a lower limit on the half-life of this process of T$^{0\nu\beta\beta}_{1/2}>$ 9$\times 10^{25}$ yr (90\% CL) in 5 years of data taking. This limit corresponds to an effective Majorana mass ranging from 55 to 133\,meV, at the top of the inverted neutrino mass hierarchy. The possibility of loading 10 times more tellurium in order to cover the majority of the inverted hierarchy region is under investigation.

Along with the \obb search, SNO+ also has the potential to measure the low energy solar neutrinos, like \textit{pep}-neutrinos. If the same purity levels as initially achieved by Borexino are reached, SNO+ can measure the \textit{pep}-neutrinos with an uncertainty less than 10\% in one year of data taking with pure liquid scintillator. Additionally, if the background is low enough SNO+ can measure CNO neutrinos.

Another physics topic that can be explored by SNO+ is the measurement of geoneutrinos in a geologically interesting location, which will be complementary to the measurements done by Borexino and KamLAND. Furthermore, SNO+ can measure reactor antineutrinos, which will help in reducing the uncertainty on the oscillation parameters. 

With its depth and low background, SNO+ has an extraordinary opportunity to measure the supernova $\nu_x$ energy spectrum for the first time. This measurement provides valuable information in order to probe and constrain supernova dynamics. Participation in SNEWS will further support a reliable early warning to the astronomical community in the event of a nearby supernova. 

During the water fill, SNO+ can search for exotic physics and set competitive limits in the invisible nucleon decay of $^{16}$O.

We expect to start operation with the water fill phase soon, followed by the liquid scintillator fill phase after a few months of data taking. The Te-loaded phase is foreseen in 2017.

\section*{Conflict of Interests}

The authors declare that there is no conflict of interests regarding the publication of this paper.

\section*{Acknowledgements}
Capital construction funds for the SNO+ experiment is provided by the Canada Foundation for Innovation (CFI). This work has been in part supported by the Science and Technology Facilities Council (STFC) of the United Kingdom (Grants no. ST/J001007/1 and ST/K001329/1), the Natural Sciences and Engineering Research Council of Canada, the Canadian Institute for Advanced Research (CIFAR), the National Science Foundation, national funds from Portugal and European Union FEDER funds through the COMPETE program, through FCT  -- Funda\c{c}\~{a}o para a Ci\^{e}ncia e a Tecnologia (Grant no. EXPL/FIS-NUC/1557/2013), the Deutsche Forschungsgemeinschaft (Grant no. ZU123/5), the European Union's Seventh Framework Programme (FP7/2007-2013, under the European Research Council (ERC) grant agreement no. 278310 and the Marie Curie grant agreement no: PIEF-GA-2009-253701), the Director, Office of Science, of the U.S. Department of Energy (Contract no. DE-AC02-05CH11231), the U.S. Department of Energy, Office of Science, Office of Nuclear Physics (Award Number DE-SC0010407), the U.S. Department of Energy (Contract No. DE-AC02-98CH10886), the National Science Foundation (Grant no. NSF-PHY-1242509) and the University of California, Berkeley. The authors acknowledge the generous support of the Vale and SNOLAB staff.

\end{document}

%% file: 201506_review_authorlist_v9b.tex
%%% A
\author[1]{\bf S.~Andringa,}
\author[2]{\bf E.~Arushanova,}
\author[3]{\bf S.~Asahi,}
\author[4]{\bf M.~Askins,}
\author[5]{\bf D.~J.~Auty,}

%%% B
\author[2,6]{\bf A.~R.~Back,}
\author[7]{\bf Z.~Barnard,}
\author[1,8]{\bf N.~Barros,}
\author[8]{\bf E.~W.~Beier,}
\author[5]{\bf A.~Bialek,}
% ...

%%%%%%%%%%%%%%%% ADDRESSES
\affil[1]{\it Laborat\'{o}rio de Instrumenta\c{c}\~{a}o e  F\'{\i}sica Experimental de Part\'{\i}culas (LIP), Av.~Elias Garcia, 14, 1$^{\circ}$,  1000-149, Lisboa, Portugal}
\affil[2]{\it Queen Mary, University of London, School of Physics and Astronomy,  327 Mile End Road, London, E1 4NS, UK}
\affil[3]{\it Queen's University, Department of Physics, Engineering Physics \& Astronomy, Kingston, ON K7L 3N6, Canada}
\affil[4]{\it University of California, 1 Shields Avenue, Davis, CA 95616, USA}
\affil[5]{\it University of Alberta, Department of Physics, 4-181 CCIS,  Edmonton, AB T6G 2E, Canada}
\affil[6]{\it University of Sussex, Physics \& Astronomy, Pevensey II, Falmer, Brighton, BN1 9QH, UK}
\affil[7]{\it Laurentian University, 935 Ramsey Lake Road, Sudbury, ON P3E 2C6, Canada}
\affil[8]{\it University of Pennsylvania, Department of Physics \& Astronomy, 209 South 33rd Street, Philadelphia, PA 19104-6396, USA}
% ...
